\documentclass[aps,pre,floatfix,twocolumn,longbibliography,nofootinbib,superscriptaddress,reprint]{revtex4-1}
\usepackage[normalem]{ulem}

\usepackage{mathptmx}

\usepackage{amssymb}
\usepackage{amsmath}

\usepackage[utf8]{inputenc}
\usepackage[T1]{fontenc}
\usepackage{t1enc}
\usepackage[english]{babel}

\usepackage{booktabs}
\usepackage{placeins}

\usepackage{graphicx} % Include figure files
\graphicspath{{img/}}

\usepackage{dcolumn}% Align table columns on decimal point
\usepackage{bm} % bold math
\usepackage{IEEEtrantools}
\usepackage{siunitx} % boldface fonts in tables

\usepackage{mathtools}
\usepackage{subfigure}

\usepackage[none]{hyphenat}

\usepackage{threeparttable}
\usepackage{enumitem}

\usepackage[usenames,dvipsnames,svgnames,table]{xcolor}
\usepackage{hyperref}
\usepackage{url}
\hypersetup{
    bookmarks=true,         % show bookmarks bar?
    unicode=false,          % non-Latin characters in Acrobat's bookmarks
    pdftoolbar=true,        % show Acrobat's toolbar?
    pdfmenubar=true,        % show Acrobat's menu?
    pdffitwindow=false,     % window fit to page when opened
    pdfstartview={FitH},    % fits the width of the page to the window
    pdftitle={},    % title
    pdfauthor={},     % author
    pdfcreator={},   % creator of the document
    pdfproducer={}, % producer of the document
    pdfkeywords={} {} {}, % list of keywords
    pdfnewwindow=true,      % links in new window
    colorlinks=true,       % false: boxed links; true: colored links
    linkcolor=red,          % color of internal links
    citecolor=Green,        % color of links to bibliography
    filecolor=magenta,      % color of file links
    urlcolor=cyan           % color of external links
}

\usepackage{algorithm}
\usepackage{algpseudocode}
\usepackage{tikz}
\newcommand*\circled[1]{\tikz[baseline=(char.base)]{
            \node[shape=circle,draw,inner sep=0.5pt] (char) {#1};}}

\makeatletter
\newcounter{phase}[algorithm]
\newlength{\phaserulewidth}
\newcommand{\setphaserulewidth}{\setlength{\phaserulewidth}}
\newcommand{\phase}[1]{%
	\vspace{-1.25ex}
	\Statex\leavevmode\llap{\rule{\dimexpr\labelwidth+\labelsep}{\phaserulewidth}}\rule{\columnwidth-15.75pt}{\phaserulewidth}
	\Statex\strut\refstepcounter{phase}\textbf{Phase~\thephase~--~#1}% Phase text
	\vspace{-1.25ex}\Statex\leavevmode\llap{\rule{\dimexpr\labelwidth+\labelsep}{\phaserulewidth}}\rule{\linewidth-15.75pt}{\phaserulewidth}}
\makeatother
\setphaserulewidth{.7pt}

\def\multiset#1#2{\left(\!\!\!\!\left({#1\atopwithdelims..#2}\right)\!\!\!\!\right)}
\def\textmultiset#1#2{\bigl(\!\!\!{\binom{#1}{#2}}\!\!\!\bigr)} % only 2 places in the paper

\newcommand{\Entropy}{S}

\newcommand{\RomanNumeralCaps}[1]{\text{\MakeUppercase{\romannumeral #1}}}
\newcommand{\bone}{\Bone}
\newcommand{\btwo}{\Btwo}
\newcommand{\bmbone}{\bib_{\RomanNumeralCaps{1}}}
\newcommand{\bmbtwo}{\bib_{\RomanNumeralCaps{2}}}
\newcommand{\Bone}{B_{\RomanNumeralCaps{1}}}
\newcommand{\Btwo}{B_{\RomanNumeralCaps{2}}}
\newcommand{\None}{N_{\RomanNumeralCaps{1}}}
\newcommand{\Ntwo}{N_{\RomanNumeralCaps{2}}}

\newcommand{\biA}{{\textbf{\textit{A}}}}
\newcommand{\bik}{{\textbf{\textit{k}}}}
\newcommand{\bie}{{\textbf{\textit{e}}}}
\newcommand{\bib}{{\textbf{\textit{b}}}}
\newcommand{\bin}{{\textbf{\textit{n}}}}

\begin{document}
\title{Community Detection in Bipartite Networks with Stochastic Blockmodels}

\author{Tzu-Chi Yen}
\email{tzuchi.yen@colorado.edu}
\affiliation{Department of Computer Science, University of Colorado, Boulder, CO, USA}

\author{Daniel B. Larremore}
\email{daniel.larremore@colorado.edu}
\affiliation{Department of Computer Science, University of Colorado, Boulder, CO, USA}
\affiliation{BioFrontiers Institute, University of Colorado, Boulder, CO, USA}


\begin{abstract}
In bipartite networks, community structures are restricted to being disassortative, in that nodes of one type are grouped according to common patterns of connection with nodes of the other type. This makes the stochastic block model (SBM), a highly flexible generative model for networks with block structure, an intuitive choice for bipartite community detection. However, typical formulations of the SBM do not make use of the special structure of bipartite networks. 
Here, we introduce a Bayesian nonparametric formulation of the SBM and a corresponding algorithm to efficiently find communities in bipartite networks which parsimoniously chooses the number of communities. 
The biSBM improves community detection results over general SBMs when data are noisy, improves the model resolution limit by a factor of $\sqrt{2}$, and expands our understanding of the complicated optimization landscape associated with community detection tasks. A direct comparison of certain terms of the prior distributions in the biSBM and a related high-resolution hierarchical SBM also reveals a counterintuitive regime of community detection problems, populated by smaller and sparser networks, where non-hierarchical models outperform their more flexible counterpart.
\end{abstract}

\maketitle
\section{Introduction}
    %%%    
    %%%   
  %%%%%%%
   %%%%%
    %%%
     %
     % Introduction
A bipartite network is defined as having two types of nodes, with edges allowed only between nodes of different types. For instance, a network in which edges connect people with the foods they eat is bipartite, as are other networks of associations between two classes of objects. Recent applications of bipartite networks include studies of plants and the pollinators that visit them~\cite{youngReconstructionPlantPollinator2019a}, stock portfolios and the assets they comprise~\cite{squartiniEnhancedCapitalassetPricing2017}, and even U.S. Supreme Court justices and the cases they vote on~\cite{guimeraJusticeBlocksPredictability2011}. More abstractly, bipartite networks also provide an alternative representation for hypergraphs in which the two types of nodes represent the hypergraph's nodes and its hyperedges, respectively~\cite{ghoshalRandomHypergraphsTheir2009,chodrowConfigurationModelsRandom2019}. 

Many networks exhibit community structure, meaning that their nodes can be divided into groups such that the nodes within each group connect to other nodes in other groups in statistically similar ways. Bipartite networks are no exception, but they exhibit a particular form of community structure because type-I nodes are defined by how they connect to type-II nodes, and vice versa. For example, in the bipartite network of people and the foods they eat, vegetarians belong to a group of nodes which are defined by the fact that they never connect to nodes in the group of meat-containing foods; meat-containing foods are defined by the fact that they never connect to vegetarians. While the group structure in this example comes from existing node categories, one can also ask whether statistically meaningful groups could be derived solely from the patterns of the edges themselves. This problem, typically called {\it community detection}, is the unsupervised task of partitioning the nodes of a network into statistically meaningful groups. In this paper, we focus on the community detection problem in bipartite networks. 

There are many ways to find community structure in bipartite networks, including both general methods---which can be applied to any network---and specialized methods derived specifically for bipartite networks. We focus on a family of models related to the stochastic blockmodel (SBM), a generative model for community structure in networks~\cite{hollandStochasticBlockmodelsFirst1983}. Since one of the SBM's parameters is a division of the nodes into groups, community detection with the SBM simply requires a method to fit the model to network data. With inference methods becoming increasingly sophisticated~\cite{peixotoBayesianStochasticBlockmodeling2018}, many variants of the SBM have been proposed, including those that accommodate overlapping communities~\cite{airoldiMixedMembershipStochastic2008,godoy-loriteAccurateScalableSocial2016}, broad degree distributions~\cite{karrerStochasticBlockmodelsCommunity2011}, multilayer networks~\cite{tarres-deulofeuTensorialBipartiteBlock2019}, hierarchical community structures~\cite{peixotoHierarchicalBlockStructures2014}, and networks with metadata~\cite{hricNetworkStructureMetadata2016,newmanStructureInferenceAnnotated2016,peelGroundTruthMetadata2017}. SBMs have also been used to estimate network structure or related observational data even if the measurement process is incomplete and erroneous~\cite{youngReconstructionPlantPollinator2019a,newmanEstimatingNetworkStructure2018,newmanNetworkStructureRich2018,peixotoReconstructingNetworksUnknown2018}. In fact, a broader class of so-called mesoscale structural inference problems, like core-periphery identification and imperfect graph coloring, can also be solved using formulations of the SBM, making it a universal representation for a broad class of problems~\cite{youngUniversalityStochasticBlock2018,olhedeNetworkHistogramsUniversality2014}.

At first glance, the existing SBM framework is readily applicable to bipartite networks. This is because, at a high level, the two types of nodes should correspond naturally to two blocks with zero edges within each block, implying that SBMs should detect the bipartite split without that split being explicitly provided. However, past work has shown that providing node type information {\it a priori} improves both the quality of partitions and the time it takes to find them~\cite{larremoreEfficientlyInferringCommunity2014}. Unfortunately those results, which relied on local search algorithms to maximize model likelihood~\cite{karrerStochasticBlockmodelsCommunity2011,larremoreEfficientlyInferringCommunity2014}, have been superseded by more recent results which show that fitting fully Bayesian SBMs using Markov chain Monte Carlo can find structures in a more efficient and non-parametric manner~\cite{peixotoNonparametricBayesianInference2017, rioloEfficientMethodEstimating2017, peixotoBayesianStochasticBlockmodeling2018}. 
These methods maximize a posterior probability, producing similar results to traditional cross validation by link predictions in many (but not all) cases~\cite{kawamotoCrossvalidationEstimateNumber2017a,valles-catalaConsistenciesInconsistenciesModel2018}. In this sense, they avoid overfitting the data, i.e., they avoid finding a large number of communities whose predictions fail to generalize.
This raises the question of whether the more sophisticated Bayesian SBM methods gain anything from being customized for bipartite networks, like the previous generation of likelihood-based methods did~\cite{larremoreEfficientlyInferringCommunity2014}.

In this paper, we begin by introducing a non-parametric Bayesian bipartite SBM (biSBM) and show that bipartite-specific adjustments to the prior distributions improve the resolution of community detection by a factor of $\sqrt{2}$, compared with the general SBM~\cite{peixotoParsimoniousModuleInference2013}. As with the general SBM, the biSBM automatically chooses the number of communities and controls model complexity by maximizing the posterior probability.

After introducing a bipartite model, we also introduce an algorithm, designed specifically for bipartite data, that efficiently fits the model to data. Importantly, this algorithm can be applied to both the biSBM and its general counterpart, allowing us to isolate both the effects of our bipartite prior distributions and the effects of the search algorithm itself. As in the maximum likelihood case~\cite{larremoreEfficientlyInferringCommunity2014}, the ability to customize the search algorithm for bipartite data provides both improved community detection results, as well as a more sophisticated understanding of the solution landscape, but unlike that previous work, this algorithm does more than simply require that blocks consist of only one type of node. Instead, the algorithm explores a two-dimensional landscape of model complexity, parameterized by the number of type-I blocks and the number of type-II blocks. This contributes to the growing body of work that explores the solution space of community detection models, including methods to sample the entire posterior~\cite{rioloEfficientMethodEstimating2017}, count of the number of metastable states~\cite{kawamotoCountingNumberMetastable2019b}, and determine the number of solution samples required to describe the landscape adequately~\cite{calatayudExploringSolutionLandscape2019a}. 

In the following sections, we introduce a degree-corrected version of the bipartite SBM~\cite{larremoreEfficientlyInferringCommunity2014}, which combines and extends two recent advances. Specifically, we recast the bipartite SBM~\cite{larremoreEfficientlyInferringCommunity2014} in a {\it microcanonical} and Bayesian framework~\cite{peixotoNonparametricBayesianInference2017} by assuming that the number of edges between groups and degree sequence are fixed exactly, instead of only in expectation. We then derive its likelihood, introduce prior distributions that are bipartite-specific, and describe an algorithm to efficiently fit the combined nonparametric Bayesian model to data. We then demonstrate the impacts of both the bipartite priors and algorithm in synthetic and real-world examples, and explore their impact on the maximum number of communities that our method can find, i.e., its resolution limit, before discussing the broader implications of this work. 

     % Introduction
     %
    %%%
   %%%%%
  %%%%%%%
    %%%
    %%%
     
\section{The microcanonical bipartite SBM}
\label{sec:micro_bisbm}
    %%%    
    %%%   
  %%%%%%%
   %%%%%
    %%%
     %
     % The microcanonical bipartite SBM
Consider a bipartite network with $\None$ nodes of type $\RomanNumeralCaps{1}$ and $\Ntwo$ nodes of type $\RomanNumeralCaps{2}$. The type-$\RomanNumeralCaps{1}$ nodes are divided into $\Bone$ blocks and the type-$\RomanNumeralCaps{2}$ nodes are divided into $\Btwo$ blocks. Let $N = \None + \Ntwo$ and $B = \Bone + \Btwo$. Rather than indexing different types of nodes separately, we index the nodes by $i = 1, 2, \dots, N$ and annotate the block assignment of node $i$ by $b_i = 1, 2, \dots, B$. A key feature of the biSBM is that each block consists of only one type of node.

Having divided nodes into blocks, we can now write down the propensities for nodes in each block to connect to nodes in the other blocks. Let $e_{rs}$ be the total number of edges between blocks $r$ and $s$. Then, let $k_i$ be the degree of node $i$.  Together, $\bie = \lbrace e_{rs} \rbrace$ and $\bik = \lbrace k_i \rbrace$ specify the degrees of each node and the patterns by which edges are placed between blocks. The number of edges attached to a group $r$ must be equal to the sum of its degrees, such that $e_r = \sum_{s} e_{rs} = \sum_{b_i = r} k_i$ for any $r$. For bipartite networks, $e_{rr} = 0$ for all $r$. We use $n_r$ to denote the number of nodes in block $r$.

Given the parameters above, one can generate a network by placing edges that satisfy the constraints imposed by $\bie$ and $\bik$. However, that network would be just one of an ensemble of potentially many networks, all of which satisfy the constraints, analogous to the configuration model~\cite{bollobasProbabilisticProofAsymptotic1980,fosdickConfiguringRandomGraph2018}. Peixoto showed how to count the number of networks in this ensemble~\cite{peixotoEntropyStochasticBlockmodel2012}, so that for a uniform distribution over that ensemble, the likelihood of observing any particular network is simply the inverse of the ensemble size. This means that, given $\bie$, $\bik$, and the group assignments $\bib = \lbrace b_i \rbrace$, computing the size of the ensemble $\|\Omega\left({\bik, \bie, \bib}\right) \|$ is tantamount to computing the likelihood of drawing a network with adjacency matrix $\biA$ from the model, $P(\biA \mid  \bik, \bie, \bib) = \|\Omega\left({\bik, \bie, \bib}\right) \|^{-1}$. Thus, treating networks as equiprobable microstates in a microcanonical ensemble leads to the microcanonical stochastic blockmodel, whose bipartite version we now develop, specifically to find communities in real-world bipartite networks. This derivation follows directly from combining the bipartite formulation of the SBM~\cite{larremoreEfficientlyInferringCommunity2014} with the microstate counting developed in~\cite{peixotoEntropyStochasticBlockmodel2012}. We introduce a new algorithm to fit the model in Sec.~\ref{sec:fitting_algm}.
 
\section{Nonparametric Bayesian SBM for Bipartite Networks}
\label{sec:nonparametric}
    %%%    
    %%%
  %%%%%%%
   %%%%%
    %%%
     %
     % Nonparametric Bayesian inference
We first formulate the community detection problem as a {\it parametric} inference procedure. The biSBM is parameterized by a partition of nodes into blocks $\bib$, the number of edges between blocks $\bie$, and the number of edges for each node, $\bik$. However, for empirical networks, we need only search the space of partitions $\bib$. This is because the microcanonical model specifies the degree sequence $\bik$ exactly, so the only way that an empirical network can be found in the microcanonical ensemble is if the parameter $\bik$ is equal to the empirically observed degree sequence. Note that, when $\bik$ and $\bib$ are both specified, $\bie$ is also exactly specified. As a consequence, community detection requires only a search over partitions of the nodes into blocks $\bib$.

In the absence of constraints on $\bib$, the maximum likelihood solution is simply for the model to memorize the data, placing each node into its own group and letting $\hat{\bie} = \biA$. To counteract this tendency to dramatically overfit, we adapt the Bayesian nonparametric framework of~\cite{peixotoNonparametricBayesianInference2017}, where the number of groups and other model parameters are determined from the data, and customize this framework for the situation in which the data are bipartite. We start by factorizing the joint distribution for the data and the parameters in this form,
\begin{equation}
    P(\biA, \bik, \bie, \bib)
	= P(\biA\mid \bik,\bie,\bib) P(\bik\mid \bie,\bib) P(\bie \mid \bib) P(\bib),
	\label{eq:joint_probability}
\end{equation}
where $P(\bik | \bie, \bib)$, $P(\bie | \bib)$, and $P( \bib )$ are prior probabilities that we will specify in later subsections. Thus, Eq.~\eqref{eq:joint_probability} defines a complete generative model for data and parameters.

The Bayesian formulation of the SBM is a powerful approach to community detection because it enables model comparison, meaning that we can use it to choose between different model classes (e.g., hierarchical vs flat) or to choose between parameterizations of the same model (e.g., to choose the number of communities).  Two approaches to model comparison, producing equivalent formulations of the problem, are useful. The first formulation is that of simply maximizing Eq.~\eqref{eq:joint_probability}, taking the view that the model which maximizes the joint probability of the model and data is, statistically most justified. The second formulation is that of minimizing the so-called {\it description length}~\cite{rissanenInformationComplexityStatistical2007}, which has a variety of interpretations (for a reviews and update, see~\cite{grunwaldMinimumDescriptionLength2007,grunwaldMinimumDescriptionLength2019}). Perhaps the most useful interpretation for our purposes is that of compression, which takes the view that the best model is one which allows us to most compress the data, while accounting for the cost to describe the model itself. In this phrasing, for a model class $M$, the description length $\Sigma_{M}\left( \biA, \bib \right)$ is given by 
$\Sigma_{M}\left( \biA, \bib \right) = -\ln P\left(\biA | \bib, M \right) - \ln P\left(\bib | M\right)$. These two terms can be interpreted as the description cost of compressing the data $\biA$ using the model and the cost of expressing the model itself, respectively. Therefore, the minimum description length (MDL) approach can be interpreted as optimizing the tradeoff between better fitting but larger models. Asymptotically, MDL is equivalent to the Bayesian Information Criterion (BIC)~\cite{schwarzEstimatingDimensionModel1978} for stochastic blockmodels under compatible prior assumptions~\cite{peixotoHierarchicalBlockStructures2014,yanModelSelectionDegreecorrected2014}. 

A complete and explicit formulation of model comparison will be provided in the context of our studies of empirical data in Sec.~\ref{sec:empirical}, using strict MDL approaches. For now, we proceed with calculating the likelihood and prior probabilities for the microcanonical biSBM and its parameters. 
     % Model comparison
     %
    %%%
   %%%%%
  %%%%%%%
    %%%
    %%%
\subsection{Likelihood for microcanonical bipartite SBM}
    %%%    
    %%%   
  %%%%%%%
   %%%%%
    %%%
     %
     % Likelihood for microcanonical bipartite SBM
The observed network $\biA$ is just one of $\|\Omega\left({\bik, \bie, \bib}\right) \|$ networks in the microcanonical ensemble which match $\lbrace \bik, \bie, \bib \rbrace$ exactly. Assuming that each configuration in the network ensemble is equiprobable, computing the likelihood is equivalent to taking the inverse of the size of the ensemble. We compute the size of the ensemble by counting the number of networks that match the desired block structure $\Omega(\bie)$ and dividing by the number of equivalent network configurations without block structure $\Xi(\biA)$, yielding, 
\begin{equation}
    P_{\text{bi}}\left( \biA \mid  \bik, \bie, \bib \right) = \|\Omega\left({\bik, \bie, \bib}\right) \|^{-1} \equiv \frac{\Xi\left( \biA \right)}{\Omega\left( \bie\right)} \ .
    \label{eq:sbm_likelihood}
\end{equation}
As detailed in Ref.~\cite{peixotoNonparametricBayesianInference2017}, the number of networks that obey the desired block structure determined by $\bie$ is given by,
\begin{equation}
    \Omega \left( \bie\right) = \frac{\prod_r{e_r!}}{\prod_{r < s} {e_{rs}!}} \ .
\end{equation}
This counting scheme assumes that half-edges are distinguishable. In other words, it differentiates between permutations of the neighbors of the same node, which are all equivalent (i.e., correspond to the same adjacency matrix). To discount equivalent permutations of neighbors, we count the number of half-edge pairings that correspond to the bipartite adjacency matrix $\biA$,
\begin{equation}
    \Xi \left( \biA \right) = \frac{\prod_i k_{i} !}{\prod_{i<j} A_{ij}!} \ .
\end{equation}
Note that while self-loops are forbidden, this formulation allows the possibility of multiedges.
     % Likelihood for microcanonical bipartite SBM
     %
    %%%
   %%%%%
  %%%%%%%
    %%%
    %%%
\subsection{Prior for the degrees}
    %%%    
    %%%   
  %%%%%%%
   %%%%%
    %%%
     %
     % Prior for the degrees        
The prior for the degree sequence follows directly from Ref.~\cite{peixotoNonparametricBayesianInference2017} because $\bik$ is conditioned on $\bie$ and $\bib$, which are bipartite. The intermediate degree distribution $\bm{\eta} = \lbrace \eta_k^r \rbrace$, with $\eta_k^r$ being the number of nodes with degree $k$ that belong to group $r$, further factorizes the conditional dependency. This allows us to write
\begin{equation}
	P\left(\bik\mid \bie, \bib \right) = P\left(\bik\mid \bm{\eta} \right) P\left(\bm{\eta} \mid  \bie,\bib \right) \ ,
	\label{eq:prior_deg}
\end{equation}
where 
\begin{equation}
    P\left(\bik\mid \bm{\eta} \right) = \prod_r \frac{\prod_k \eta_k^r!}{n_r!}
\end{equation}
is a uniform distribution of degree sequences constrained by the overall degree counts, and 
\begin{equation}
	P\left(\bm{\eta} \mid  \bie,\bib \right)
		= \prod_r q(e_r, n_r)^{-1}
	\label{eq:eta_hyperprior}
\end{equation}
is the distribution of the overall degree counts. The quantity $q\left( m, n\right)$ is the number of restricted partitions of the integer $m$ into at most $n$ parts~\cite{andrewsTheoryPartitions1998}. It can be computed via the following recurrence relation, 
\begin{equation}
\label{eqn:int_part_exact}
	q\left( m, n\right)	= q\left( m, n-1\right) + q\left(m-n, n\right),
\end{equation}
with boundary conditions $q\left( m, 1\right) = 1$ for $m > 0$, and $q\left(m, n\right) = 0$ for $m \leq 0$ or $n \leq 0$. With this, computing $q\left( m, n\right)$ for $m \leq M$ and $n \leq m$ requires $\mathcal{O}( M^2)$ additions of integers. In practice, we precompute $q(m, n)$ using the exact Eq.~\eqref{eqn:int_part_exact} for $m \leq 10^4$ (or $m \leq E$ when the network is smaller), and resort to approximations~\cite{peixotoNonparametricBayesianInference2017} only for larger arguments.

For sufficiently many nodes in each group, the hyperprior Eq.~\eqref{eq:eta_hyperprior} will be overwhelmed by the likelihood, and the distribution of Eq.~\eqref{eq:prior_deg} will approach the actual degree sequence. In such cases, the prior and hyperprior naturally learn the true degree distribution, making them applicable to heterogeneous degrees present in real-world networks.
     % Prior for the degrees
     %
    %%%
   %%%%%	
  %%%%%%%
    %%%
    %%%
\subsection{Prior for the node partition}
    %%%    
    %%%   
  %%%%%%%
   %%%%%
    %%%
     %
     % Prior for the node partition
The prior for the partitions $\bib$ also follows Ref.~\cite{peixotoNonparametricBayesianInference2017} in its general outline, but the details require modification for bipartite networks. We write the prior for $\bib$ as the following Bayesian hierarchy
\begin{equation}
	P_{\text{bi}}\left( \bib \right) = P\left( \bib\mid\bin \right) P\left( \bin\mid  B \right)P\left( B \right) \ , \label{eq:partitionprior}
\end{equation}
where $\bin = \lbrace n_r \rbrace$, the number of nodes in each group. We then assume that this prior can be factorized into independent priors for the partitions of each type of node, i.e., $P_{\text{bi}}\left( \bib \right) = P\left( \bmbone \right) P\left( \bmbtwo \right)$. This allows us to treat the terms of Eq.~\eqref{eq:partitionprior} as
\begin{equation}
    P\left( \bib\mid \bin \right) = 
    \left ( \frac{\prod_{\substack{\text{type-I}\\ \text{groups } r}} n_r!}{N_\text{I}!} \right ) 
    \left ( \frac{\prod_{\substack{\text{type-II}\\ \text{groups } s}} n_s!}{N_\text{II}!} \right ) \ ,
    \label{eq:partitionprior_1}
\end{equation} 
\begin{equation}
	P\left( \bin\mid  B \right) = \binom{\None - 1}{\Bone - 1}^{-1} \binom{\Ntwo - 1}{\Btwo - 1}^{-1}\ ,
	    \label{eq:partitionprior_2}
\end{equation}
and
\begin{equation}
	P\left(B\right) = \None^{-1}\Ntwo^{-1}\ .
	    \label{eq:partitionprior_3}
\end{equation}
Equation~\eqref{eq:partitionprior_2} is a uniform hyperprior over all such histograms on the node counts~$\bin$, while Eq.~\eqref{eq:partitionprior_3} is a prior for the number of nonempty groups itself. This Bayesian hierarchy over partitions accommodates heterogeneous group sizes, allowing it to model the  group sizes possible in real-world networks.
     % Prior for the node partition
     %
    %%%
   %%%%%
  %%%%%%%
    %%%
    %%%
\subsection{Prior for the bipartite edge counts}
    %%%    
    %%%   
  %%%%%%%
   %%%%%
    %%%
     %
     % Prior for the bipartite edge counts
We now introduce the prior for edge counts between groups, $\bie$, which also requires modification for bipartite networks. While the edge count prior for general networks is parameterized by the number of groups $B$, the analogous prior for bipartite networks is parameterized by $\Bone$ and $\Btwo$. We therefore modify the counting scheme of Ref.~\cite{peixotoNonparametricBayesianInference2017}, written for general networks, to avoid counting non-bipartite partitions that place edges between nodes of the same type. Our prior for edge counts between groups is therefore 
\begin{equation}
	P_{\text{bi}}\left( \bie \mid \bib \right)	= \multiset{\Bone \Btwo}{E}^{-1}\ ,
	\label{eq:uniform_bipartite_prior}
\end{equation}
where $\Bone \Btwo$ counts the number of group-to-group combinations when edges are allowed only between type-$\RomanNumeralCaps{1}$ and type-$\RomanNumeralCaps{2}$ nodes. The notation $\textmultiset{\Bone \Btwo}{E} = \binom{\Bone \Btwo+E-1}{E}$ counts the number of histograms with $\Bone \Btwo$ bins whose counts sum to~$E$. Similar to the uniform prior for general networks~\cite{peixotoNonparametricBayesianInference2017}, it is unbiased and maximally non-informative, but by neglecting mixed-type partitions, this prior results in a more parsimonious description.  In later sections, we show that this modified formulation enables the detection of smaller blocks, improving the so-called resolution limit, by reducing model complexity for larger $\Bone$ and $\Btwo$.
     % Prior for the bipartite edge counts
     %
    %%%
   %%%%%
  %%%%%%%
    %%%
    %%%
\subsection{Model summary}
    %%%    
    %%%   
  %%%%%%%
   %%%%%
    %%%
     %
     % Model summary
Having fully specified the priors in previous subsections, we now substitute our calculations into Eq.~\eqref{eq:joint_probability}, the joint distribution for the biSBM, yielding, 
\begin{widetext}
\begin{align}
	P_{\text{bi}}\left(\biA,\! \bik,\! \bie,\! \bib\right)
	\!=\! \frac{\prod_i{k_i!}\prod_{r<s}{e_{rs}!}}{\prod_r{e_{r}!}\prod_{i<j}{A_{ij}!}} 
	\prod_{r}{\frac{\prod_k{\eta_k^r!}}{n_r!}\frac{1}{q\left(e_r, n_r\right)}} 
 \multiset{\Bone \Btwo}{E}^{\!-1}
 \frac{\prod_r{n_r!}}{\None! \Ntwo!} \binom{\None\!-\!1}{\Bone\!-\!1}^{\!-1} \binom{\Ntwo\!-\!1}{\Btwo\!-\!1}^{\!-1} \frac{1}{\None \Ntwo}\ .
\label{eq:full_posterior}
\end{align}
\end{widetext} 

Inference of the biSBM reduces to the task of sampling this distribution efficiently and correctly. Although Eq.~\eqref{eq:full_posterior} is somewhat daunting, note that $\bik$ and $\bie$ are implicit functions of the partition $\bib$, meaning Eq.~\eqref{eq:full_posterior} depends only on the data and the partition $\bib$. This opens the door to efficient sampling of the posterior distribution via Markov chain Monte Carlo which we discuss in Sec.~\ref{sec:fitting_algm}. 

\subsection{Comparison with the hierarchical SBM}

In deriving the biSBM, we replaced the SBM's uniform prior for edge counts with a bipartite formulation Eq.~\eqref{eq:uniform_bipartite_prior}. However, one can instead replace it with a Bayesian hierarchy of models~(Eq.~\eqref{eq:hi_edge_count}; \cite{peixotoHierarchicalBlockStructures2014}). In this hierarchical SBM (hSBM), the matrix $\bie$ is itself considered as an adjacency matrix of a multigraph with $B$ nodes and $E$ edges, allowing it to be modeled by a second SBM. Of course, the second SBM also has an edge count matrix with the same number of edges and fewer nodes, so the process of modeling each edge count matrix using another SBM can be done recursively until the model has only one block. In so doing, the hSBM typically achieves a higher posterior probability (which corresponds to higher compression, from a description length point of view) than non-hierarchical (or ``flat'') models, and can therefore identify finer-scale community structure.

The hSBM's edge count prior allows it to find finer scale communities and more efficiently represent network data. However, as we will see, when the network is small and has no hierarchical structure, the hSBM can actually underfit the data, finding too few communities, due to the overhead of specifying a hierarchy even when none exists. The scenarios in which the flat bipartite prior has advantages over its hierarchical counterpart are explored in Sec.~\ref{sec:benchmark}. 
% We supply the derivation of the prior for the hierarchical biSBM in Appendix~\ref{appendix:hSBM}.

     % Model summary
     %
    %%%
   %%%%%
  %%%%%%%
    %%%
    %%%
     % The microcanonical bipartite SBM    
     %
    %%%
   %%%%%
  %%%%%%%
    %%%
    %%%        
\section{Fitting the model to data}
\label{sec:fitting_algm}

The mathematical formulation of the biSBM takes full advantage of a network's bipartite structure to arrive at a better model. Here, we again make use of that bipartite structure to accelerate and improve our ability to fit the model, Eq.~\eqref{eq:full_posterior}, to network data. 

At a high level, our algorithm for model fitting consists of two key routines. The first routine is typical of SBM inference, and uses Markov chain Monte Carlo importance sampling~\cite{metropolisEquationStateCalculations1953,hastingsMonteCarloSampling1970,peixotoEfficientMonteCarlo2014}, followed by simulated annealing, to explore the space of partitions, conditioned on fixed community counts. In this routine, we accelerate mixing time by making use of the bipartite constraint, specifying a Markov chain only over states (partitions) with one type of node in each block. Importantly, this constraint has the added effect that we must fix both block counts, $\Bone$ and $\Btwo$, separately. 

The second routine of our algorithm consists of an adaptive search over the two-dimensional space of possible $(\Bone, \Btwo)$, using the ideas of dynamic programming~\cite{cormenIntroductionAlgorithms3rd2009,ericksonAlgorithms2019}. It attempts to move quickly through those parts of the $(\Bone, \Btwo)$ plane that are low probability under Eq.~\eqref{eq:full_posterior} without calling the MCMC routine, and instead allocating computation time for the regions that better explain the data. The result is an effective algorithm, with two separable routines, which makes full use of the network's bipartite structure, allowing us to either maximize or sample from the posterior Eq.~\eqref{eq:full_posterior}.

One advantage of having decoupled routines in this way is that the the partitioning engine is a modular component which can be swapped out for a more efficient alternative, should one be engineered or discovered. Reference implementations of two SBM partitioning algorithms, a Kernighan-Lin-inspired local search~\cite{kernighanEfficientHeuristicProcedure1970,karrerStochasticBlockmodelsCommunity2011,larremoreEfficientlyInferringCommunity2014} and the MCMC algorithm, are freely available as part of the~\texttt{bipartiteSBM} library~\cite{yenBipartiteSBMPythonLibrary}.

Alternative methods for model fitting exist. For instance, it is possible to formulate a Markov chain over the entire space of partitions whose stationary distribution is the full posterior, without conditioning on the number of groups. In such a scheme, transitions in the Markov chain can create or destroy groups~\cite{rioloEfficientMethodEstimating2017}, and the Metropolis-Hastings principles guarantee that this chain will eventually mix. However, this approach turns out to be too slow to be practical because the chain gets trapped in metastable states, extending mixing times. 

Another alternative approach is to avoid our two-dimensional search over $\Bone$ and $\Btwo$, and instead search over $B = \Bone + \Btwo$.  This is the approach of Ref.~\cite{peixotoParsimoniousModuleInference2013}, where, after proving the existence of an optimal number of blocks $B$, a golden-ratio one-dimensional search is used to efficiently find it.

\subsection{Inference routine}
\label{sec:inference_algm}
    %%%    
    %%%   
  %%%%%%%
   %%%%%
    %%%
     %
     % Inference algorithm
The task of the MCMC inference routine is to maximize Eq.~\eqref{eq:full_posterior}, conditioned on fixed values of $\Bone$ and $\Btwo$. Starting from an initial partition $\bib_{\text{init}}$, the MCMC algorithm explores the space of partitions with fixed $\Bone$ and $\Btwo$ by proposing changes to the block memberships $\bib$, and then accepting or rejecting those moves with carefully specified probabilities. As is typical, those probabilities are chosen so that the probability that the algorithm is at any particular partition is equal to the posterior probability of that partition, given $\Bone$ and $\Btwo$, by enforcing the Metropolis-Hastings criterion.

Rather than initializing the MCMC procedure from a fully random initial partition, we instead use an agglomerative initialization~\cite{peixotoHierarchicalBlockStructures2014} which reduces burn-in time and avoids getting trapped in metastable states that are common when group sizes are large. The agglomerative initialization amounts to putting each node in its own group and then greedily merging pairs of groups of matching types until the specified $\Bone$ and $\Btwo$ remain. 

After initialization, each step consists of proposing to move a node $i$ from its current group $r$ to a new group $s$. Following~\cite{peixotoNonparametricBayesianInference2017}, proposal moves are generated efficiently in a two-step procedure. First, we sample a random neighbor $j$ of node $i$ and inspect its group membership $b_j$. Then, with probability $\epsilon B / ( e_{b_j} + \epsilon B )$ we choose $s$ uniformly at random from $\lbrace 1, 2, \dots, B \rbrace$; otherwise, we choose $s$ with probability proportional to the number of edges leading to that group from group $b_j$, i.e., proportional to $e_{b_j s}$. 

A proposed move which would violate the bipartite structure by mixing node types, or which would leave group $r$ empty, is rejected with probability one. A valid proposed move is accepted with probability
\begin{equation}
    a = \min \left \{ 1, \frac{p\left(b_i = s \rightarrow r \right)}{p\left(b_i = r \rightarrow s \right)} \exp{ \left(- \beta \Delta \Entropy \right)} \right \} \ ,
    \label{eq:acceptance_probability}
\end{equation}
where 
\begin{equation}
  	p\left(b_i = r \rightarrow s \right) = \sum_{t}{R_t^i} \frac{e_{ts} + \epsilon}{e_t + \epsilon B} \ .
  	\label{eq:proposal_moves}
\end{equation}
Here, $R_t^i$ is the fraction of neighbors of node $i$ which belong to block $t$ and and $\epsilon > 0$ is an arbitrary parameter that enforces ergodicity. The term $\beta$ is an inverse-temperature parameter, and $\Delta \Entropy$ is the difference between the entropies of the biSBM's microcanonical ensemble in its current state and in its proposed new state. With this in mind, 
\begin{equation}
	\Delta \Entropy = \Entropy|_{b_i = s} - \Entropy|_{b_i = r} =  \ln \frac{P\left( \biA, \bik, \bie, \bib \right)}{P\left( \biA', \bik', \bie', \bib' \right)} \ ,
	\label{eq:delta_entropy}
\end{equation}
where variables without primes represent the current state ($b_i =r$) and variables with primes correspond to the state being proposed ($b_i = s$).  

The initialization, proposal, and evaluation steps of the algorithm above are fast. With continuous bookkeeping of the incident edges to each group, proposals can be made in time $\mathcal{O}\left(k_i\right)$, and are engineered to substantially improve the mixing times since they remove an explicit dependency on the number of groups which would otherwise be present with the fully random moves~\cite{peixotoHierarchicalBlockStructures2014}. Then, when evaluating Eq.~\eqref{eq:delta_entropy}, we need only a number of terms proportional to $k_i$. In combination, the cost of an entire ``sweep,'' consisting of one proposed move for each node in the network, is $\mathcal{O}\left(E\right)$. The overall number of steps necessary for MCMC inference is therefore $\mathcal{O}\left( \tau E \right)$, where $\tau$ is the average mixing time of the Markov chain, independent of $B$. 

Our \texttt{bipartiteSBM} implementation~\cite{yenBipartiteSBMPythonLibrary} has the following default settings, chosen to stochastically maximize Eq.~\eqref{eq:full_posterior} for fixed $\Bone$ and $\Btwo$ via a simulated annealing process. We first let $\epsilon = 1$, and perform $10^3$ sweeps at $\beta=1$ to reach equilibrated partitions. Then we perform zero-temperature ($\beta \rightarrow \infty$) sweeps, in which only moves leading to a strictly lower entropy are allowed. We keep track of the system's entropy during this process and exit the MCMC routine when no record-breaking event is observed within a $2 \times 10^3$ sweeps window, or when the number of sweeps exceeds $10^4$, whichever is earlier. The partition $\bib$ at the end corresponds to the lowest entropy. Equivalently stated, this partition $\bib$ corresponds to the minimum description length or highest posterior probability, for fixed $\Bone$ and $\Btwo$. The minimal entropy at each stage is bookmarked for future decision-making processes.

The bipartite MCMC formulation is more than just similar to its general counterpart. In fact, one can show that for fixed $\Bone$ and $\Btwo$, the Markov chain transition probabilities dictated by Eq.~\eqref{eq:delta_entropy} are identical for the uniform bipartite edge count prior Eq.~\eqref{eq:uniform_bipartite_prior} and its general equivalent introduced in~\cite{peixotoNonparametricBayesianInference2017}. This means that the MCMC algorithm explores the same entropic landscape for both bipartite and general networks when $\Bone$ and $\Btwo$ are fixed. As we will demonstrate in Sec.~\ref{sec:benchmark}, however, by combining the MCMC routine with both the novel search routine over the block counts and the more sensitive biSBM priors, we can better infer model parameters in bipartite networks.
     % Inference algorithm
     %
    %%%
   %%%%%
  %%%%%%%
    %%%
    %%%
\begin{figure}[tp]
\centering
\includegraphics[width=0.88\linewidth]{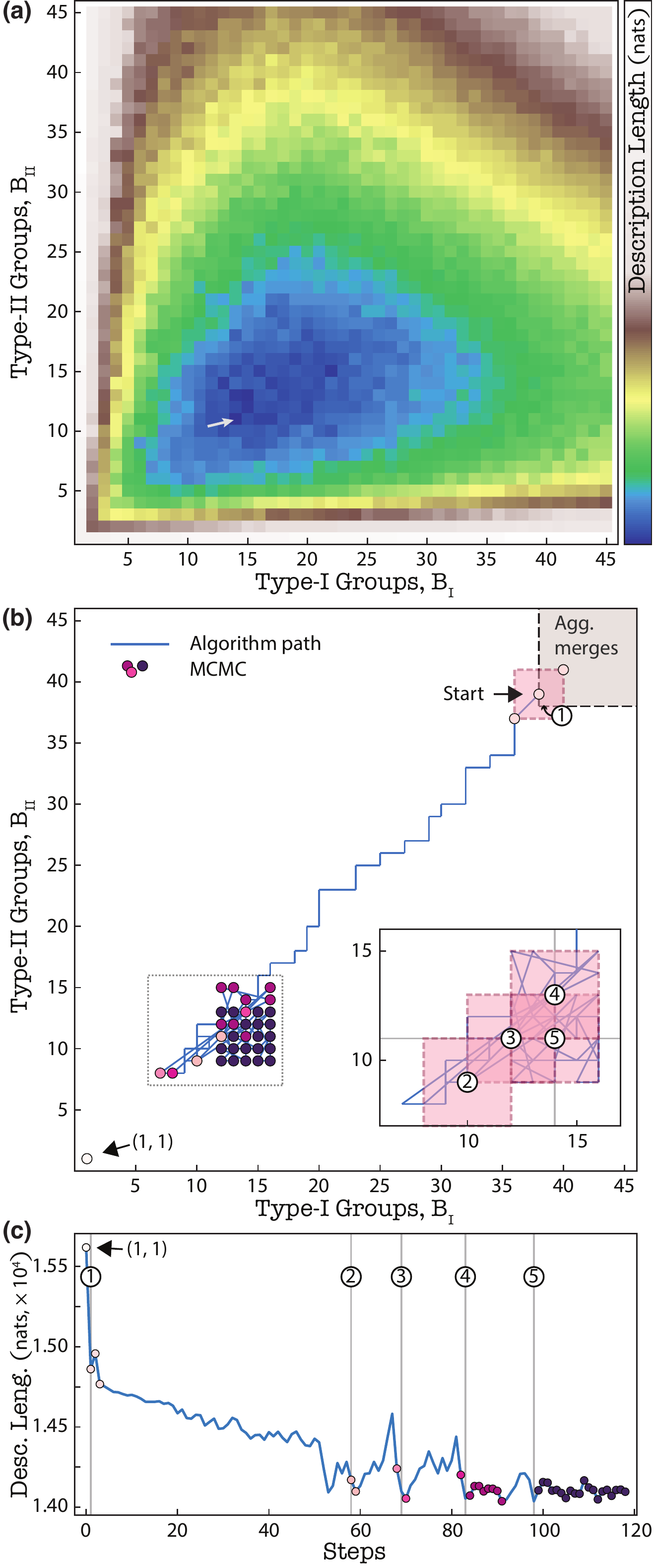}
\caption[]{Diagram showing the biSBM community detection algorithm on the description length landscape of the malaria gene-substring network~\cite{larremoreNetworkApproachAnalyzing2013}. 
(a)~Each square in the heatmap shows the result of fitting a model using MCMC at the specified $\left (\Bone, \Btwo\right )$. The color bar scales linearly. An arrow indicates the minimizing point. (b)~Trajectory of the efficient search routine over the landscape shown in the top panel. Circles indicate where MCMC inference was required. Pink shaded regions show neighborhoods of exhaustive local search, with sequential order indicated by \circled{1} to \circled{5}. (c)~Change of description length values as the algorithm progresses. Shaded circles show the steps at which the 36 MCMC calculations were performed. The minimizing point at $(11,14)$ was found during local search \circled{4} and confirmed during local search \circled{5}.}
\label{fig:heuristic}
\end{figure}    
\subsection{Search routine}
\label{sec:search_algm}
    %%%    
    %%%   
  %%%%%%%
   %%%%%
    %%%
     %
     % Search algorithm

The task of the search routine is to maximize Eq.~\eqref{eq:full_posterior} over the $\left( \Bone, \Btwo\right)$ plane, i.e., to find the optimal number of groups. However, maximizing Eq.~\eqref{eq:full_posterior} for any fixed choice of $\left( \Bone, \Btwo\right)$ requires the MCMC inference introduced above, motivating the need for an efficient search. If we were to treat the network as unipartite, a one-dimensional convex optimization on the total number of groups $B = \Bone + \Btwo$ with a search cost of $\mathcal{O}\left( \ln N \right)$~\cite{peixotoParsimoniousModuleInference2013} could be used. On the other hand, exhaustively exploring the plane of possibilities would incur a search cost of $\mathcal{O}(B_\text{max}^2)$, where $B_\text{max}$ is the maximum value of $B$ which can be detected. In fact, our experiments indicate that neither the general unipartite approach nor the naive bipartite approach is optimal. The plane search is too slow, while the line search undersamples local maxima of the $\left( \Bone, \Btwo\right)$ landscape, which is typically multimodal. Instead, we present a recursive routine that runs much faster than exhaustive search, which parameterizes the tradeoff between search speed and search accuracy by rapidly finding the high-probability region of the $\left( \Bone, \Btwo\right)$ plane without too many calls to the more expensive MCMC routine.

We provide only a brief outline of the search algorithm here, supplying full details in Appendix~\ref{appendix:algorithm}. The search is initialized with each node in its own block. Blocks are rapidly agglomerated until $\min\left( \Bone, \Btwo \right) = \lfloor \sqrt{2 E}/2 \rfloor$. This is the so-called {\it resolution limit}, the maximum number of communities that our algorithm can reliably find, which we discuss in detail in Sec.~\ref{sec:resolution_limit}. Equation~\eqref{eq:full_posterior} will never be maximized prior to reaching this frontier. During this initial phase, we also compute the posterior probability of the trivial bipartite partition with $(1,1)$ blocks, as a reference for the next phase.

\begin{figure*}[!t]
\centering
  \begin{center}
    \includegraphics[width=0.9\linewidth]{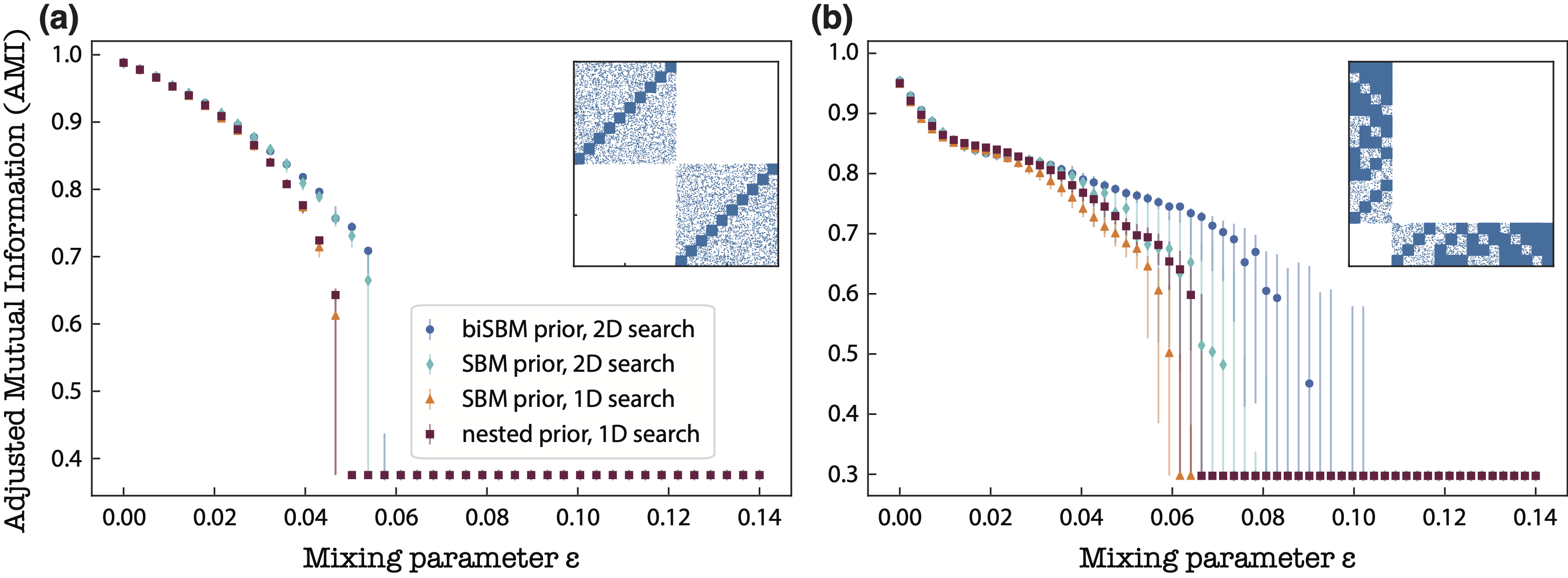}
    \caption[]{Numerical tests of the recovery of planted structure in synthetic networks with $N=10^4$ nodes. Each point shows the median of $10^2$ replicates of the indicated model and algorithm (see legend) and error bars show $25\%-75\%$ quantiles. Insets show the structure of the problems at moderate $\epsilon$. (a) A test meant to be easy: mean degree $5$, equally sized groups, and $\Bone = \Btwo = 10$. (b) A test meant to be challenging: mean degree $15$, equally sized groups, and $\Bone = 4$ and $\Btwo = 15$.}
    \label{fig:benchmark}
  \end{center}
\end{figure*}

Next, we search the region of the ($\Bone$,$\Btwo$) plane within the resolution frontier to find a local maximum of Eq.~\eqref{eq:full_posterior} by adaptively reducing the number of communities. In this context, a local maximum is defined as an MCMC-derived partition with exactly ($\Bone$,$\Btwo$) blocks, whose posterior probability is larger than the posterior probabilities for MCMC-derived partitions at nearby values ($\Bone\pm h$,$\Btwo\pm h$), for a chosen neighborhood size $h$. From the initial partition at the resolution frontier, we merge blocks, selected greedily from a stochastically sampled set of proposed merges. Here, because the posterior probability is a tiny value, it is computationally more convenient to work with the model entropy $\Entropy$, which is related to the posterior probability by $\Entropy = - \ln P$.  Proposed merges are evaluated by their entropy after merging, but without calling the MCMC routine to optimize the post-merge partition. Because MCMC finds better (or no worse) fits to the data, this means that these post-merge entropies are approximate upper bounds of the best-fit entropy, given the post-merge number of blocks. We therefore use this approximate upper bound to make the search adaptive: whenever a merge would produce an upper-bound approximation that is a factor $1+\Delta_0$ higher than the current best $\Entropy$, a full MCMC search is initialized at the current grid point. Otherwise, merges proceed rapidly since the approximate entropy is extremely cheap to compute. Throughout this process, the value of $\Delta_0$ is estimated from the data to balance accuracy and efficiency, and it adaptively decreases as the search progresses (Appendix~\ref{appendix:algorithm}). The algorithm exits when it finds a local minimum on the entropic landscape, returning the best overall partition explored during the search.
  
In practice, a typical call to the algorithm takes the form of (i) a rapid agglomerative merging phase from $(\None,\Ntwo)$ blocks to the resolution limit frontier; (ii) many agglomerative merges to move along candidate local minima that rely on approximated entropy; (iii) more deliberate and MCMC-reliant neighborhood searches to examine candidate local minima. These phases are shown in Fig.~\ref{fig:heuristic}. The algorithm has total complexity $\mathcal{O}(m h^2)$, where $m$ is the number of times that an exhaustive neighborhood search is performed. When $h=2$, we find $m < 3$ for most empirical networks examined. This algorithm is not guaranteed to find the global optimum, but due to the typical structure of the $\left( \Bone, \Btwo\right)$ optimization landscape for bipartite networks, we have found it to perform well for many synthetic and empirical networks, and it tends consistently estimate the number of groups (see Sec.~\ref{sec:resolution_limit}). An implementation is available in the \texttt{bipartiteSBM} library~\cite{yenBipartiteSBMPythonLibrary}.
     % Search algorithm
     %
    %%%
   %%%%%
  %%%%%%%
    %%%
    %%%
\section{Reconstruction performance}\label{sec:benchmark}
    %%%    
    %%%   
  %%%%%%%
   %%%%%
    %%%
     %
     % Reconstruction performance        
In this section, we examine our method's ability to correctly recover the block structure in synthetic bipartite networks where known structure has been intentionally hidden. In each test, we begin by creating a bipartite network with unambiguous block structure, and then gradually mix that structure with noise until the planted blocks disappears entirely, creating a sequence of community detection problems that are increasingly challenging~\cite{mooreComputerSciencePhysics2017}. The performance of a community detection method can then be measured by how well it recovers the known partition over this sequence of challenges. 

The typical synthetic test for unipartite networks is the {\it planted partition model}~\cite{condonAlgorithmsGraphPartitioning2001} in which groups have $\omega_{rr} = \omega_\text{in}$ assortative edges, and $\omega_{rs} = \omega_\text{out}$ disassortative edges for $r \neq s$. When the total expected degree for each group is fixed, the parameter $\epsilon = \omega_\text{out}/\omega_\text{in}$ controls the ambiguity of the planted blocks. Unambiguous assortative structure corresponds to $\epsilon=0$ while $\epsilon=1$ corresponds to a fully random graph. Here, we consider a straightforward translation of this model to bipartite networks in which the nodes are again divided into blocks according to a planted partition. As in the unipartite planted partition model, non-zero entries of the block affinity matrix take on one of two values but due to the fact that all edges are disassortative, we replace $\omega_\text{in}$ and $\omega_\text{out}$ with $\omega_{\text{+}}$ or $\omega_{\text{–}}$ to avoid confusion (see insets of Fig.~\ref{fig:benchmark}). By analogy, we let $\epsilon=\omega_{\text{-}}/\omega_{\text{+}}$ while fixing the total expected degree for each group, so that $\epsilon=0$ corresponds to highly resolved communities which blend into noise as $\epsilon$ grows. 

We present two synthetic tests using this bipartite planted partition model, designed to be easy and difficult, respectively. In the easy test, the unambiguous structure consists of $\None\!=\!\Ntwo\!=\!\tfrac{1}{2}10^4$ nodes, divided evenly into $\Bone\!=\!\Btwo\!=\!10$ blocks of 500 nodes each, with a mean degree $\langle k \rangle = 5$. Each type-I block is matched with a type-II block so that the noise-free network consists of exactly 10 bipartite components, with zero edges placed between nodes in different components by definition. In the hard test, the unambiguous structure consists of $N\!=\!10^4$ nodes divided evenly into $\Bone=4$ and $\Btwo=15$ blocks of approximately equal size, with mean degree $\langle k \rangle=15$. The relationships between the groups in the hard test are more complex, so the insets of Fig.~\ref{fig:benchmark} provide schematics of the adjacency matrices of both tests under a moderate amount of noise. In both cases, node degrees were drawn from a power-law distribution with exponent $\alpha=2$, and for a fixed $\epsilon$, networks were drawn from the canonical degree-corrected stochastic blockmodel~\cite{karrerStochasticBlockmodelsCommunity2011,larremoreEfficientlyInferringCommunity2014}.

We test four methods' abilities to recover the bipartite planted partitions, in combinations that allow us to separate the effects of using our bipartite {\it model} (Sec.~\ref{sec:nonparametric}) and our bipartite {\it search algorithm} (Sec.~\ref{sec:fitting_algm}), in comparison to existing methods. The first method maximizes the biSBM posterior using our 2D search algorithm. The second method keeps the 2D search algorithm, but examines the effects of the bipartite-specific edge count prior by replacing it with the general SBM's edge count prior [i.e., replacing Eq.~\eqref{eq:uniform_bipartite_prior} with Eq.~\eqref{eq:uniform_prior}]. The third method uses the same general SBM edge count prior as the second, but uses a 1D bisection search~\cite{peixotoParsimoniousModuleInference2013} to examine the effects of the 2D  search. The fourth method maximizes the hierarchical SBM posterior using a 1D bisection search. For the first two cases, we use our  \texttt{bipartiteSBM} library~\cite{yenBipartiteSBMPythonLibrary}, while for the latter two, we use the \texttt{graph-tool} library~\cite{peixotoGraphtoolPythonLibrary2014}.  In all cases, we enforce type-specific MCMC move proposals to avoid mixed-type groups. 

In the easy test, we find that the bipartite search algorithm introduced in Sec.~\ref{sec:fitting_algm} performs better than the one-dimensional searches (Fig.~\ref{fig:benchmark}a). Because the one-dimensional search algorithm assumes that the optimization landscape is unimodal, we reasoned that other modes may emerge as $\epsilon$ increases. To test this, we generated networks within the transition region ($\epsilon \approx 0.054$) and then conducted an exhaustive survey of plausible $(\Bone,\Btwo)$ values using MCMC with the general SBM. This revealed two basins of attraction, located at $(8, 8)$ and $(1,1)$, explaining the SBM's performance. This bimodal landscape can therefore hinder search in one dimension by too quickly attracting the algorithm to the trivial bipartite partition. Perhaps surprisingly then, a similar exhaustive survey of the $(\Bone,\Btwo)$ plane using the bipartite model revealed that near the transition $\epsilon$, the biSBM has a local optimum with {\it more than} the planted $(10,10)$ blocks.

In the hard case, we find that it is not the bipartite search that enables the biSBM to outperform the other methods, but rather the bipartite posterior (Fig.~\ref{fig:benchmark}b). An exploration of the outputs of the general searches shows that when they fail, they tend to find an incorrect number of blocks, which should total $19$ [corresponding to the planted $(4,15)$ blocks]. To understand this failure mode in more detail, we fixed $B=19$ and used MCMC to fit the general SBM~\cite{peixotoGraphtoolPythonLibrary2014}. This led to solutions in which $\Bone \approx \Btwo$, revealing that the performance degradation, relative to the biSBM, was due to a tendency for that particular algorithmic implementation of the SBM to find more balanced numbers of groups. Interestingly, near their respective transitions values of $\epsilon$, both the SBM and biSBM tend to find more groups than were planted in the hard test, thus overfitting the data. To explore this further, we again conducted exhaustive surveys of the $(\Bone,\Btwo)$ plane using MCMC and found that under both models, the posterior surfaces are consistently multimodal, with attractive peaks corresponding to more communities than the planted $(4,15)$. However, only the bipartite search algorithm introduced in Sec.~\ref{sec:fitting_algm} finds overfitted partitions with too many groups; the unipartite search algorithms instead return underfitted models with too few groups, balanced between the node types. 

In sum, our synthetic network tests reveal two phenomena. First, the biSBM with bipartite search is able to extract structure from higher levels of noise than the alternatives, making it an attractive option for bipartite community detection with real data. However, our tests also reveal that the posterior surfaces of both the SBM and biSBM degenerate in unexpected ways near the detectability transition~\cite{decelleAsymptoticAnalysisStochastic2011,mosselReconstructionEstimationPlanted2015,kawamotoDetectabilityThresholdsGeneral2017,ricci-tersenghiTypologyPhaseTransitions2019a}.

     % Reconstruction performance
     %
    %%%
   %%%%%
  %%%%%%%
    %%%
    %%% 
\section{Resolution Limit}\label{sec:resolution_limit}
    %%%    
    %%%   
  %%%%%%%
   %%%%%
    %%%
     %
     % Resolution limit
\begin{figure}[!t]
	\centering
	\includegraphics[width=0.9\linewidth]{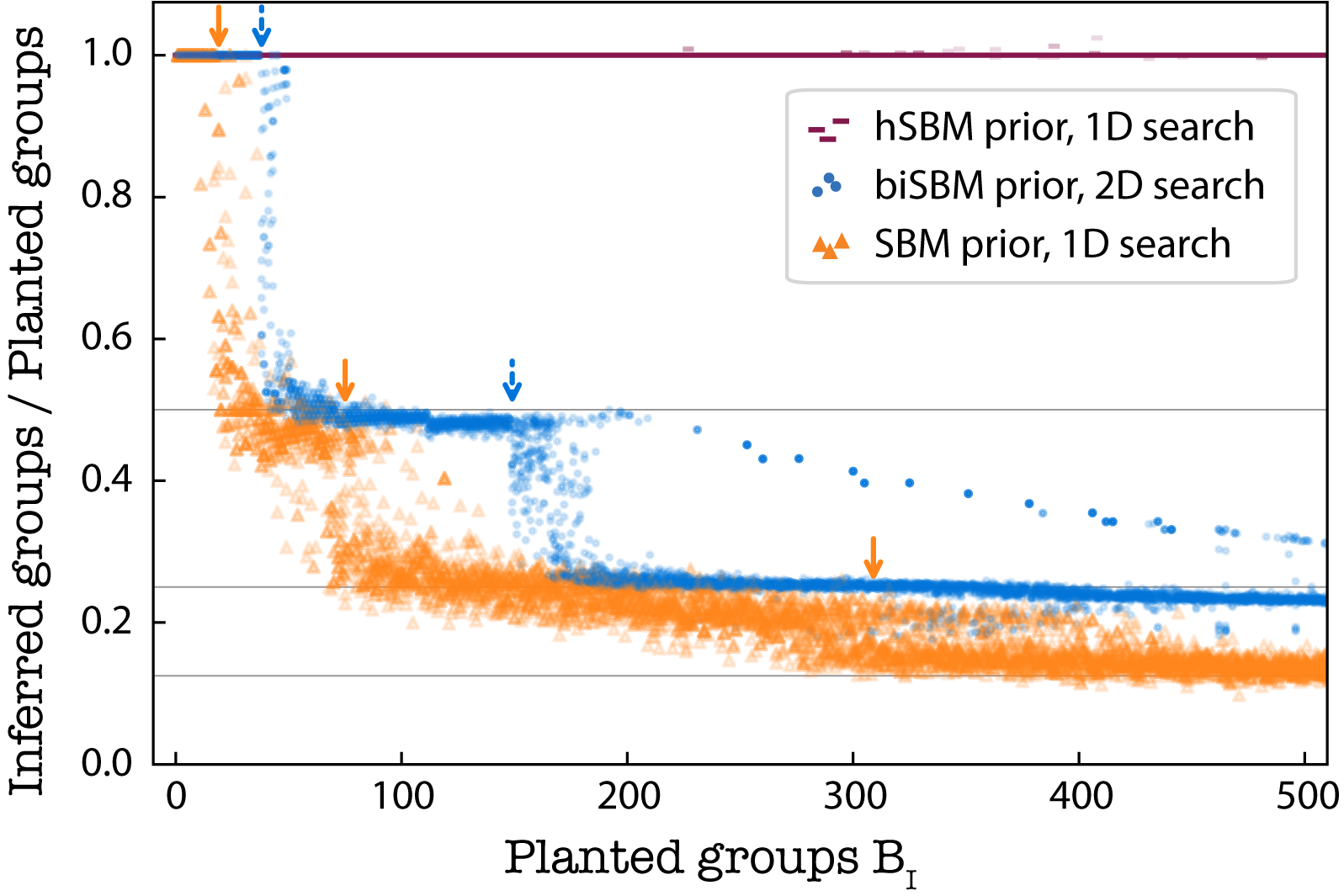}
	\caption[]{A numerical experiment on bipartite cliques to demonstrate the resolution limit. As an increasing number of bipartite cliques with $10$ nodes of each type are presented to the SBM, biSBM, and hSBM (see legend), the hSBM continues to find all cliques while the SBM and biSBM begin to merge pairs, quartets, and eventually octets of cliques. Arrows indicate analytical predictions of merge transitions from posterior odds ratios, with colors matching the legend. Note that biSBM transitions occur at twice the value of $B$ as SBM transitions, showing the biSBM's expanded resolution limit.}
	\label{fig:resolution} 
\end{figure}     
Community detection algorithms exhibit a {\it resolution limit}, an upper bound on the number of blocks that can be resolved in data, even when those blocks are seemingly unambiguous. For instance, using the general SBM, only $B_\text{max} = \mathcal{O} \left(N^{1/2}\right)$ groups can be detected~\cite{peixotoNonparametricBayesianInference2017}, while the higher resolution of the hierarchical SBM improves this scaling to $B_\text{max} = \mathcal{O} \left( {N}/\ln{N}\right)$~\cite{peixotoHierarchicalBlockStructures2014}. In this section we investigate the resolution limit of the biSBM numerically and analytically.

Our numerical experiment considers a network of $\Bone = \Btwo = \tilde{B}$ bipartite cliques of equal size, with $10$ nodes of each type per biclique and therefore $100$ edges per biclique. To this network, we repeatedly apply the SBM, the hSBM, and biSBM, and record the number of blocks found each time, varying $\tilde{B}$ between $1$ and $510$. For small values of $\tilde{B}$, all three algorithms infer $\tilde{B}$ blocks, but as the number of blocks increases, solutions which merge pairs, then quartets, and then octets become favored (Fig.~\ref{fig:resolution}). The hSBM continues to find $\tilde{B}$ blocks, as expected. 

The exact value of $\tilde{B}$ at which merging blocks into pairs becomes more attractive can be derived by asking when the corresponding posterior odds ratio, comparing a model with $\tilde{B}$ bicliques to a model with $\tilde{B}/2$ biclique pairs, exceeds one, 
\begin{equation}
	\Lambda(\tilde{B}) = \frac{P(\biA,\bik,\bie_\text{clique pairs},\bib_\text{clique pairs})}{P(\biA,\bik,\bie_\text{cliques},\bib_\text{cliques})}\ .
	\label{eq:bayes_resolution}
\end{equation}
When there are $10$ nodes of each type per biclique and $100$ edges, $\Lambda(\tilde{B})$ exceeds 1 when $\tilde{B}=19$ for the SBM and $\tilde{B}=38$ for the biSBM (Fig.~\ref{fig:resolution}; arrows). A similar calculation predicts the transition from biclique pairs to biclique quartets at $\tilde{B}=75$ for the SBM and $\tilde{B}=149$ for the biSBM (Fig.~\ref{fig:resolution}; arrows). Numerical experiments confirm these analytical predictions, but noisily, due to the stochastic search algorithms involved, and the fact the optimization landscapes are truly multimodal, particularly near points of transition.

\begin{figure}[!t]
\centering
\includegraphics[width=0.9\linewidth]{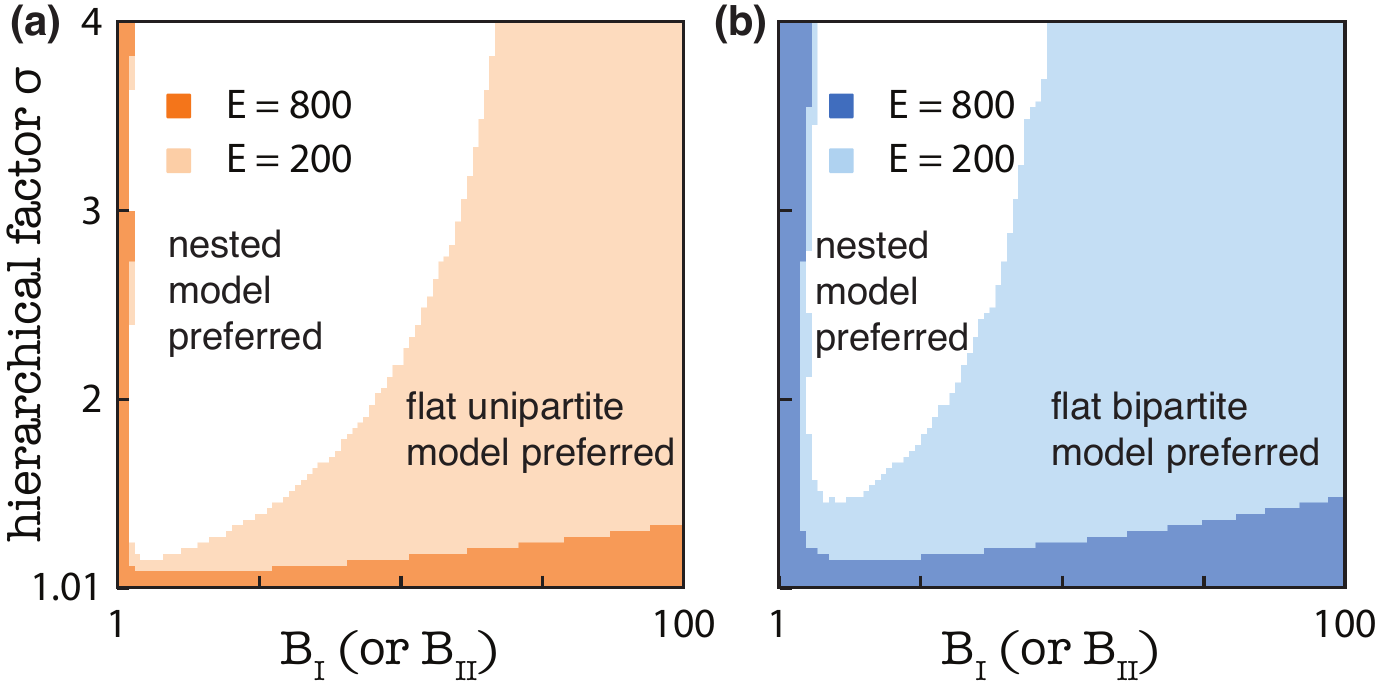}
\caption[]{Comparison of the description lengths resulting from prior distribution over edge counts using the biSBM, SBM, and hSBM priors. Regions where a flat prior has a lower description length than the hierarchical prior are shaded for (a) the SBM and (b) the biSBM. Flat priors are favored when there are fewer edges, more groups, and a smaller hierarchical branching factor $\sigma$ (defined in Sec.~\ref{sec:resolution_limit}). The flat-model regime is larger for the biSBM than the SBM, as described in Sec.~\ref{sec:resolution_limit}.} 
\label{fig:edge_count_prior}
\end{figure}

The posterior odds ratio calculations above can be generalized, and show that the biSBM extends the resolution transitions twice as far as the SBM for the transitions from $B\! \to\! \tfrac{1}{2}B \! \to\! \tfrac{1}{4}B\! \to\! \dots$, and so on, but still undergoes the same transitions eventually. Thus, both models exhibit the same resolution limit scaling $B_\text{max} = \mathcal{O} \left(N^{1/2}\right)$, but with resolution degradations that occur at $N$ for the SBM occurring at $2N$ for the biSBM.  Therefore, the resolution limit of the biSBM is $\sqrt{2}$ larger than the SBM for the same number of nodes. One can alternatively retrace the analysis of Ref.~\cite{peixotoNonparametricBayesianInference2017}, but for the biSBM applied to bicliques to derive the same $\sqrt{2}$ resolution improvement.

This constant-factor improvement in resolution limit may seem irrelevant, given that the major contribution of the hierarchical SBM was to change the order of the limit to $B_\text{max} = \mathcal{O} \left( {N}/\ln{N}\right)$~\cite{peixotoHierarchicalBlockStructures2014}. However, we find that, on the contrary, the $\sqrt{2}$ factor improvement for the biSBM expands a previously uninvestigated regime in which flat models outperform their hierarchical cousin. When given the biclique data, the hSBM finds a hierarchical division where at each level $l$, the number of groups decreases by a factor $\sigma_l$, except at the highest level where it finds a bipartite division. Assuming that $\sigma_l = \sigma$, we have $B_l = 2 \tilde{B}_l$, where $B_l = \tilde{B}/\sigma^{l-1}$. The hSBM's prior for edge counts Eq.~\eqref{eq:hi_edge_count} can be factored into uniform distributions over multigraphs at lower levels and over an SBM at the topmost level, leading to,
\begin{IEEEeqnarray}{rCl}\IEEEeqnarraymulticol{3}{l}
    {P_\text{lower}\left( \bie \right) = \prod_{l=1}^{\log_\sigma{\tilde{B}}}{\multiset{\sigma^2}{E\sigma^{l}/\tilde{B}}}^{-\tilde{B}/\sigma^{l}}}
    \nonumber\\& \times &
    \frac{{\sigma!}^{2\tilde{B}/\sigma^{l}}}{\left( \tilde{B}/\sigma^{l-1}\right)!^2}\binom{\tilde{B}/\sigma^{l-1}-1}{\tilde{B}/\sigma^{l}-1}^{-2} ,\nonumber\\
    \label{eq:biclique_lower_multigraph}
\end{IEEEeqnarray}
and,
\begin{equation}
    P_\text{topmost}\left( \bie \right) = \multiset{\textmultiset{2}{2}}{E}^{-1} \ .
    \label{eq:biclique_topmost_multigraph}
\end{equation}
By comparing $P_\text{hier} = P_\text{lower} P_\text{topmost}$ with the corresponding terms from the biSBM [Eq.~\eqref{eq:uniform_bipartite_prior}] or the corresponding equation for the SBM [Eq.~\eqref{eq:uniform_prior}], we can identify regimes in which a flat model better describes network data than the nested model.

\begin{table*}[ht]
\caption{Results for 24 empirical networks. Number of nodes $n_{\RomanNumeralCaps{1}}$, $n_{\RomanNumeralCaps{2}}$, mean degree $\langle k \rangle$, number of type-$\RomanNumeralCaps{1}$ groups $\Bone$, and number of type-$\RomanNumeralCaps{2}$ groups $\Btwo$, and description length per edge $\Sigma/E$. Superscripts: b-biSBM, g-SBM, h-hSBM. $L$ indicates the number of levels found by the hSBM. Reported values indicate best of 100 independent runs. Unless otherwise noted, data are accessible from the Colorado Index of Complex Networks (ICON)~\cite{clausetColoradoIndexComplex}. The confidence level is marked with asterisks$^\text{a}$.}
\centering
 \begin{tabular}%
 	{
 	% 10 columns
 	>{\raggedright\arraybackslash}p{5.1cm} %
	>{\raggedleft\arraybackslash}p{1.1cm} %
	>{\raggedleft\arraybackslash}p{1.1cm} %
	>{\raggedleft\arraybackslash}p{1.1cm} %
	>{\centering\arraybackslash}p{1.6cm} %
	>{\centering\arraybackslash}p{1.6cm} %
	>{\centering\arraybackslash}p{1.6cm} %
	>{\raggedright\arraybackslash}p{1.3cm} %
	>{\raggedright\arraybackslash}p{1.0cm} %
	>{\raggedright\arraybackslash}p{1.0cm} %
	 }
 \hline\hline
 Dataset & $\None$ & $\Ntwo$ & $\langle k \rangle$ & $(\Bone^{\text{b}}, \Btwo^{\text{b}})$ & $(\Bone^{\text{g}}, \Btwo^{\text{g}})$ & $(\Bone^{\text{h}}, \Btwo^{\text{h}})$ & $\langle L + 1 \rangle$ & $\Sigma^\text{b}/E$ & $\Sigma^\text{h}/E$\\ [0.3ex]
 \hline
 Southern women interactions~\cite{jonesDeepSouthSocial1942} & 18 & 14 & 5.56 & (1, 1) & (1, 1) & (1, 1) & 2.0 & \bfseries2.15$^*$ & 2.26 \\
 Joern plant-herbivore web~\cite{joernFeedingPatternsGrasshoppers1979} & 22 & 52 & 4.97 & (2, 2) & (1, 1) & (1, 1) & 2.0 & \bfseries2.64$^*$ & 2.74 \\ 
 Swingers and parties~\cite{niekampSexualAffiliationNetwork2013} & 57 & 39 & 4.83 & (1, 1) & (1, 1) & (1, 1) & 2.0 & \bfseries2.92$^*$ & 2.97 \\
 McMullen pollination web~\cite{mcmullenFlowervisitingInsectsGalapagos1993} & 54 & 105 & 2.57 & (2, 2) & (2, 2) & (1, 1) & 2.0 & \bfseries2.87$^*$ & 3.02 \\
 Ndrangheta criminals~\cite{dimilanoOrdinanzaDiApplicazione2011} & 156 & 47 & 4.48 & (3, 4) & (3, 3) & (3, 4) & 2.87 & \bfseries3.44$^*$ & 3.49 \\
 Abu Sayyaf kidnappings$^\text{b}$~\cite{gerdesAssessingAbuSayyaf2014} & 246 & 105 & 2.28 & (2, 2) & (1, 1) & (1, 1) & 2.0 & \bfseries4.50$^*$ & 4.54\\ 
 Virus-host interactome~\cite{rozenblatt-rosenInterpretingCancerGenomes2012} & 53 & 307 & 2.52 & (2, 2) & (1, 1) & (1, 1) & 2.0 & \bfseries3.78$^*$ & 3.81\\
 Clements-Long plant-pollinator~\cite{clementsExperimentalPollinationOutline1923} & 275 & 96 & 4.98 & (1, 1) & (1, 1) & (1, 1) & 2.0 & \bfseries3.45$^*$ & 3.47 \\ 
 Human musculoskeletal system~\cite{murphyStructureFunctionControl2018} & 173 & 270 & 4.30 & (7, 8) & (5, 5) & (8, 8) & 4.01 & \bfseries3.94 & 3.94 \\ 
 Mexican drug trafficking$^\text{b}$~\cite{cosciaKnowingWhereHow2012} & 765 & 10 & 16.1 & (12, 8) & (8, 7) & (10, 6) & 3.11 & \bfseries1.26$^*$ & 1.29 \\
 Country-language network~\cite{kunegisKONECTKoblenzNetwork2013} & 254 & 614 & 2.89 & (4, 5) & (2, 2) & (4, 3) & 2.11 & \bfseries 4.53$^*$ & 4.56 \\ 
 Malaria gene similarity~\cite{larremoreNetworkApproachAnalyzing2013} & 297 & 806 & 5.38 & (15, 16) & (6, 6) & (25, 20) & 4.95 & 4.73 & \bfseries4.67$^*$ \\
 Protein complex-drug~\cite{nacherModularityProteinComplex2012} & 739 & 680 & 5.20 & (20, 22) & (14, 14) & (35, 39) & 5.06 & 3.65 & \bfseries3.50$^{**}$ \\ 
 Robertson plant-pollinator~\cite{robertsonFlowersInsectsLists1928} & 456 & 1428 & 16.2 & (20, 18) & (11, 11) & (20, 19) & 4.0 & \bfseries3.10$^*$ & 3.10 \\ 
 Human gene-disease network~\cite{gohHumanDiseaseNetwork2007} & 1419 & 516 & 4.06 & (13, 14) & (9, 9) & (35, 36) & 5.04 & 5.02 & \bfseries4.80$^{**}$ \\ 
 Food ingredients-flavors web~\cite{ahnFlavorNetworkPrinciples2011} & 1525 & 1107 & 27.9 & (27, 69) & (20, 29) & (42, 130) & 4.91 & 2.55  & \bfseries 2.51$^{**}$\\  
 Wikipedia doc-word network~\cite{gerlachNetworkApproachTopic2018} & 63 & 3140 & 24.8 & (22, 206) & (18, 23) & (29, 71) & 4.17 & 1.58 & \bfseries 1.51$^{**}$\\ 
 Foursquare check-ins~\cite{yangFinegrainedPreferenceawareLocation2013} & 2060 & 2876 & 11.0 & (65, 66) & (40, 40) & (244, 248) & 5.2 & 5.92 & \bfseries5.09$^{**}$\\ 
 Ancient metabolic network~\cite{goldfordRemnantsAncientMetabolism2017} & 5651 & 5252 & 4.22 & (18, 22) & (5, 5) & (17, 21) & 4.26 & \bfseries5.68$^{**}$ & 5.82\\
 Marvel Universe characters~\cite{alberichMarvelUniverseLooks2002} & 6486 & 12942 & 9.95 & (68, 72) & (67, 62) & (365, 314) & 6.24 & 4.70 & \bfseries4.42$^{***}$ \\
 Reuters news stories~\cite{lewisRCV1NewBenchmark2004} & 19757 & 38677 & 33.5 & (396, 440) & (87, 108) & (294, 463) & 6.25 & 4.22 & \bfseries4.16$^{***}$ \\ 
 IMDb movie-actor dataset$^\text{c}$ & 53158 & 39768 & 6.49 & (91, 92) & (69, 68) & (264, 265) & 6.22 & 7.40 & \bfseries7.30$^{***}$ \\ 
 YouTube group memberships~\cite{misloveMeasurementAnalysisOnline2007} & 94238 & 30087 & 4.72 & (62, 66) & (37, 38) & (221, 238) & 5.9 & \bfseries7.07$^{**}$ & 7.13 \\  
 DBpedia writer network~\cite{auerDBpediaNucleusWeb2007} & 89355 & 46213 & 2.13 & (22, 26) & (2, 3) & (2, 3) & 2.16 & \bfseries10.32$^{**}$ & 10.41\\
 [0.5ex] 
 \hline\hline
 \end{tabular}
 \begin{tablenotes}[flushleft]
 \small
 \item $^\text{a}$ Via the posterior odds ratio: $^{*}: \Lambda < 10^{-2}$;\quad $^{**}: \Lambda < 10^{-100}$;\quad $^{***}: \Lambda < 10^{-10000}$.
 \item $^\text{b}$ Temporal data with timestamps are aggregated, making a multigraph.
 \item $^\text{c}$ Data available at \url{https://www.imdb.com/interfaces}. IMDb copyright permits redistribution of data only in unaltered form.
 \end{tablenotes}
 \label{table:benchmark}
\end{table*}

Figure~\ref{fig:edge_count_prior} shows regimes in which the flat model is preferred for both the SBM and biSBM. These regimes are larger for the biSBM than the SBM, as expected, and are larger when the hierarchical branching factor $\sigma$ decreases---indeed, if the data are less hierarchical, the hierarchical model is expected to have less of an advantage. The flat-model description is also favored when there are fewer edges and more groups, suggesting that in order for the nested model to be useful, it requires sufficient data to support its more costly nested architecture. A number of real-world networks that fall into this flat-model regime are described in the following section. We note that our definition of this regime relies on assumptions of perfect inference and a fixed branching factor at each level of the hSBM's hierarchy. These assumptions may not always hold. 
     % Resolution limit
     %
    %%%
   %%%%%
  %%%%%%%
    %%%
    %%%

\begin{figure*}[t]
\centering
\includegraphics[width=0.9\linewidth]{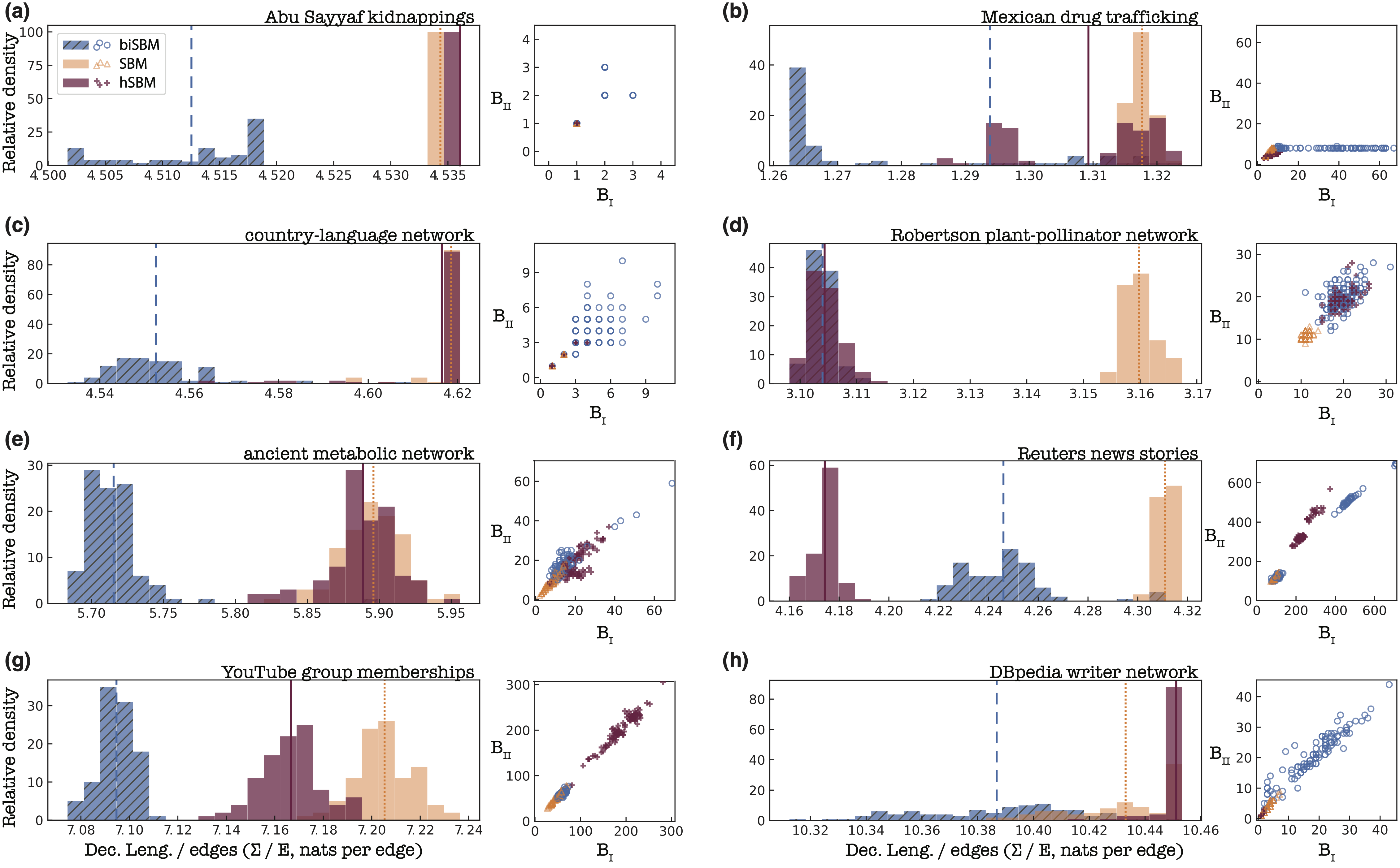}
\caption[]{Repeated application of models (see legend in panel a) with default algorithms produces distributions of the description length and the number of groups, for eight of the empirical networks listed in Table~\ref{table:benchmark}. Vertical lines mark the value of the mean description length.}
\label{fig:empirical}
\end{figure*}
\section{Empirical networks}\label{sec:empirical}
    %%%    
    %%%   
  %%%%%%%
   %%%%%
    %%%
     %
     % Empirical networks
We now examine the application of the biSBM to a corpus of real-world networks ranging in size from $N=32$ to $N=135,568$ nodes, across social, biological, linguistic, and technological domains. While it was typical of past studies to measure a community detection method by its ability to recapitulate known metadata labels, we acknowledge that this approach is inadvisable for a number of theoretical and practical reasons~\cite{peelGroundTruthMetadata2017} and instead compare the biSBM to the SBM and hSBM using Bayesian model selection. 

In general, to compare one partition-model pair $\left( \bib_0, M_0 \right)$ and an alternative pair $\left( \bib_1, M_1 \right)$, we can compute the posterior odds ratio, 
\begin{equation}
    \Lambda = \frac{P\left( \bib_0, M_0 | \biA \right)}{P\left( \bib_1, M_1 | \biA \right)}	= \frac{P\left(\biA, \bib_0 | M_0 \right)}{P\left(\biA, \bib_1 | M_1 \right)} \times \frac{P\left( M_0\right)}{P\left( M_1\right)} \ .
\end{equation}
Model $\left( \bib_0, M_0 \right)$ is favored when $\Lambda > 1$ and model $\left( \bib_1, M_1 \right)$ is favored when $\Lambda < 1$, with the magnitude of difference from $\Lambda=1$ indicating the degree of confidence in model selection~\cite{jeffreysTheoryProbability1998}. In the absence of any {\it a priori} preference for either model, $P\left(M_0\right) = P\left(M_1\right)$, meaning that the ratio of probabilities $\Lambda$ can be alternatively expressed via the difference in description lengths, $\Lambda \equiv \exp({\Sigma_1 - \Sigma_0})$. [Recall that the description length $\Sigma_\ell$ for the combined model $(\bib_\ell,M_\ell)$ and data $\biA$ can be written as the negative log of the posterior probability, as introduced in Sec.~\ref{sec:nonparametric}.] In what follows, we compare the hSBM to the biSBM and without loss of generality choose $M_1$ to be whichever model is favored so that $\Lambda$ simply expresses the magnitude of the odds ratio. Note that by construction, the biSBM always outperforms the flat SBM.

As predicted in the previous section, the biSBM's flat prior is better when networks are  smaller and sparser, while for larger networks the hSBM generally performs better by building a hierarchy that results in a more parsimonious model (Table~\ref{table:benchmark}). Indeed, the majority of larger networks are better described using the hSBM (Table~\ref{table:benchmark}; rightmost columns), but exceptions do exist, including the ancient metabolic network~\cite{goldfordRemnantsAncientMetabolism2017}, YouTube memberships~\cite{misloveMeasurementAnalysisOnline2007}, and DBpedia writer network~\cite{auerDBpediaNucleusWeb2007}, which share the common feature of low density. The Robertson plant-pollinator network~\cite{robertsonFlowersInsectsLists1928}, on the other hand, is neither small nor particularly sparse, and yet the biSBM is still weakly preferred over the hSBM. 

Differences between models, based only on their {\it maximum a posteriori} (i.e., minimum description length) estimates, may overlook additional complexity in the models' full posterior distributions. We repeatedly sample from the posterior distributions of the SBM, biSBM, and hSBM for $8$ networks from Table~\ref{table:benchmark}, showing both posterior description length distributions and inferred block count distributions  (Fig.~\ref{fig:empirical}). Generally, all three models exhibit similar description-length variation, but due to the 2D search introduced in Sec.~\ref{sec:fitting_algm}, the biSBM returns partitions with wider variation in $\Bone$ and $\Btwo$. For instance, the drug trafficking network~\cite{cosciaKnowingWhereHow2012}, a multigraph with $\None \gg \Ntwo$, has a bimodal distribution of description lengths under the hSBM, while the biSBM finds plausible partitions for a wide variety of $\Bone$ values (Fig.~\ref{fig:empirical}b). On the other hand, posterior distributions for the country-language network~\cite{kunegisKONECTKoblenzNetwork2013} are all unimodal, but the biSBM finds probable states with wide variation in description length and block counts, while the hSBM samples from a small region (Fig.~\ref{fig:empirical}c). This can happen when the network is small, since the hSBM requires sufficiently complicated data to justify a hierarchy, while the biSBM finds a variety of lower description length partitions. In fact, viewing the same datasets through the lenses of these different models' priors can quite clearly shift the location of posterior peaks. This is most clearly visible in the Reuters network~\cite{lewisRCV1NewBenchmark2004}, for which the models have unambiguous and non-overlapping preferred states (Fig.~\ref{fig:empirical}f). 

Briefly, we note that model comparison is possible here due to the fact that all of the models we considered are SBMs with clearly specified posterior distributions. Broader comparisons between community detection models of entirely different classes are also possible, for which we suggest Ref.~\cite{ghasemianEvaluatingOverfitUnderfit2019}.
     % Empirical networks
     %
    %%%
   %%%%%
  %%%%%%%
    %%%
    %%%
\section{Discussion}\label{sec:discussion}
    %%%    
    %%%   
  %%%%%%%
   %%%%%
    %%%
     %
     % Discussion     
This paper presented a bipartite microcanonical stochastic blockmodel (biSBM) and an algorithm to fit the model to network data. Our work is built on two foundations, developing a bipartite SBM~\cite{larremoreEfficientlyInferringCommunity2014} with a more sophisticated microstate counting approach~\cite{peixotoEntropyStochasticBlockmodel2012}. The model itself follows in the footsteps of Bayesian SBMs~\cite{peixotoHierarchicalBlockStructures2014,peixotoNonparametricBayesianInference2017} but with key modifications to the prior distribution and the search algorithm that more correctly account for the fact that some partitions are strictly prohibited when a network is bipartite. As a result, the biSBM is able to resolve community structure in bipartite networks better than the general SBM, demonstrated in tests with synthetic networks (Fig.~\ref{fig:benchmark}).

The resolution limit of the biSBM is greater than the general SBM by a factor of $\sqrt{2}$. We demonstrated this mathematically and in a simple biclique-finding test (Fig.~\ref{fig:resolution}). This analysis led us to directly compare the priors for the biSBM and the hierarchical SBM, which hinted at an unexpected regime in which the biSBM provides a better model than the hSBM. This regime, populated by smaller, sparser, and less hierarchical networks, was found in real data where model selection favored the biSBM (Table~\ref{table:benchmark}). 

\begin{figure}[!t]
\centering
\includegraphics[width=0.9\linewidth]{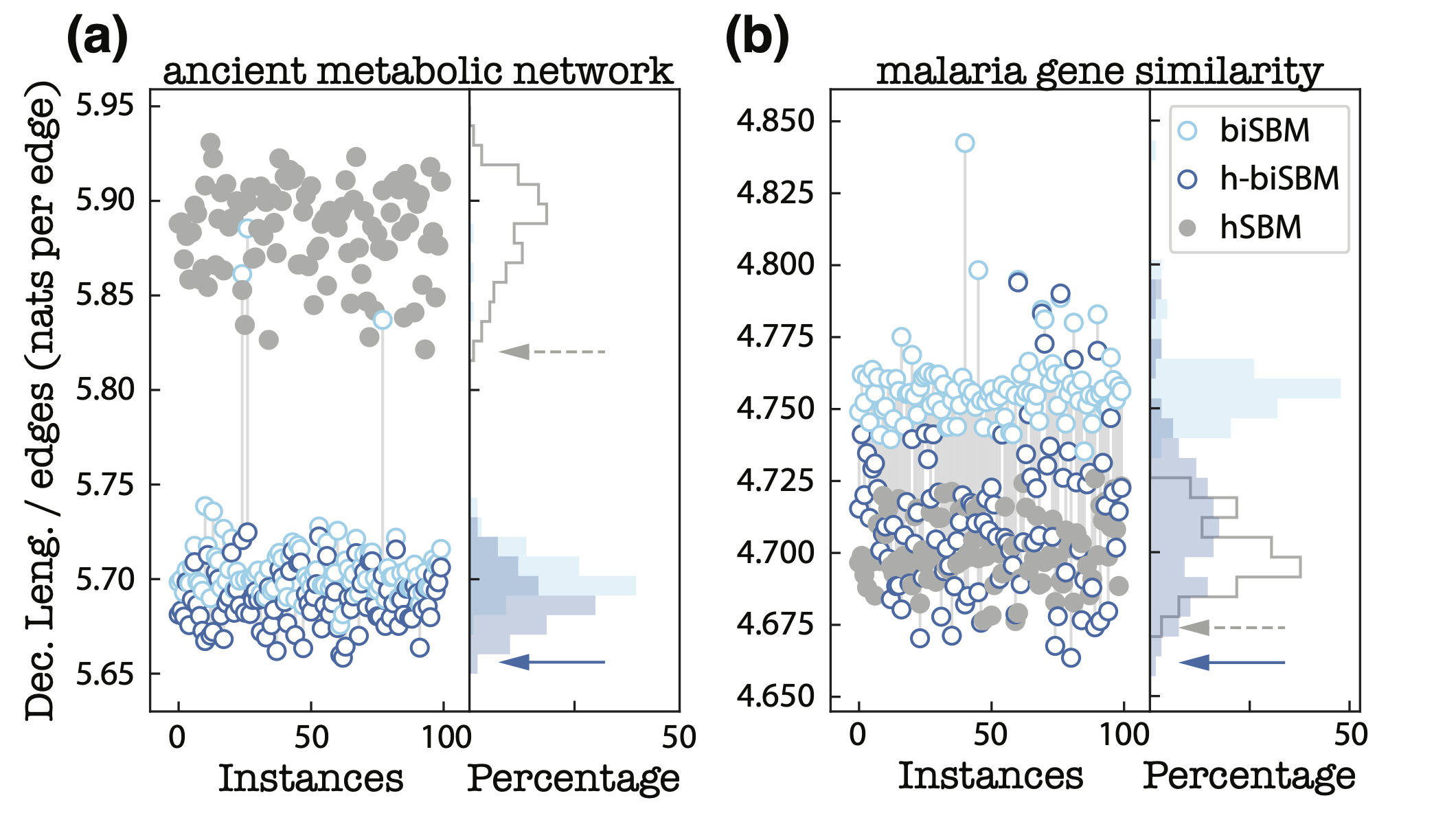}
\caption[]{Scatter plots and histograms of description length for the ancient metabolic (a;~\cite{goldfordRemnantsAncientMetabolism2017})~and malaria gene similarity (b;~\cite{larremoreNetworkApproachAnalyzing2013})~networks from 100 independent experiments. Grey vertical lines connect biSBM results with their matching h-biSBM hierarchical results. Arrows in histograms mark the MDL points from the hSBM (grey) and by the h-biSBM (blue).}
\label{fig:metabolic}
\end{figure}    

How should we understand these networks that are better described by our flat model than a hierarchical one? One possibility is that these networks are simply ``flat'' and so any hierarchical description simply wastes description-length bits on a model which is too complex. Another possibility is that this result can be explained not by the mathematics of the models but by the algorithms used to fit the models. In fact, our tests with synthetic networks show clear differences between models {\it and} algorithms, with the 2D search algorithm introduced here providing better fits to data than a 1D search (Fig.~\ref{fig:benchmark}). However, this finding alone does not actually differentiate between the two possible explanations, and so we constructed the following simple test. 

To probe the differences between the biSBM and hSBM as {\it models} vs differences in their model-fitting {\it algorithms}, we combined both approaches in a two-step protocol: Fit the biSBM to network data and then build an optimal hierarchical model upon that fixed biSBM base. Unless the data are completely flat, this hierarchy-building process will further reduce the description length, providing a more parsimonious model. If the hybrid h-biSBM provides a superior description length to the hSBM, our observations can be attributed to differences in model-fitting algorithms. In fact, this is precisely what we find. 

Figure~\ref{fig:metabolic} shows repeated application of the biSBM, hSBM, and hybrid h-biSBM to the ancient metabolic network~\cite{goldfordRemnantsAncientMetabolism2017} and the malaria genes network~\cite{larremoreNetworkApproachAnalyzing2013}. In the ancient metabolic network, the biSBM already outperformed the hSBM, so the hybrid model results in only marginal improvements in description length. However, doing so also creates hierarchies with an average depth of $\langle L \rangle = 3.85$ layers, compared with the $\langle L \rangle = 3.27$ layers found by hSBM natively. In other words, we can achieve a deeper hierarchy in addition to a more parsimonious model when using the flat biSBM partition at the lowest level. This suggests that, in fact, not all of the hSBM's underperformance can be attributed to the ancient metabolic network's being ``flat,'' since a hierarchy can be constructed upon the biSBM's inferred structure. In the malaria genes network, although the hSBM outperformed the biSBM, the hybrid model was superior to both. Since the hybrid partitions are, in principle, available to the hSBM, our conclusion is that the 2D search algorithm we presented is actually finding better partitions. Put another way, there are further opportunities to improve the depth and speed of algorithms to fit stochastic blockmodels to real-world data, particularly when bipartite or other structure in the data can be exploited. 

Finally, this work shows how both models and algorithms can reflect the structural constraints of real-world network data, and how doing so improves model quality. While our work addresses only community detection for bipartite networks, generalizations of both the mathematics and search algorithms could in principle be derived for multi-partite networks in which more complicated rules exist for how node types are allowed to connect.

\acknowledgements
The authors thank Tiago Peixoto, Tatsuro Kawamoto, Pan Zhang, Joshua Grochow, and Jean-Gabriel Young for stimulating discussions. DBL was supported in part by the Santa Fe Institute Omidyar Fellowship. The authors thank the BioFrontiers Institute at the University of Colorado Boulder and the Santa Fe Institute for the use of their computational facilities.     
     % Discussion
     %
    %%%
   %%%%%
  %%%%%%%
    %%%
    %%%
    
\appendix
\setcounter{figure}{0} \renewcommand{\thefigure}{A.\arabic{figure}}  
\renewcommand{\figurename}{Algorithm}
    %%%    
    %%%   
  %%%%%%%
   %%%%%
    %%%
     %
     % Appendix A    
\section{Recursive 2D search algorithm}\label{appendix:algorithm}

In this appendix, we elaborate on the recursive search algorithm sketched in Sec.~\ref{sec:search_algm}. Our overarching problem is to find the $(\Bone, \Btwo)$ pair that minimizes the description length. We use dynamic programming to solve this problem efficiently, observing that it has the following two properties.

(1) {\it Optimal substructure.}---If we collectively inspect the solutions that lead to local minima, then the best of those determines the global minimum.

(2) {\it Overlapping subproblems.}---To verify the existence of a local minimum, we have to compute the description length for its neighborhood points. There are many subproblems which are solved again and again and their solutions can be stored in a table so that these need not to be recomputed.

Our recursive algorithm is summarized in Algms.~\ref{alg:main} and~\ref{alg:helper}. Due to its recursive construction, a base case (or smallest subproblem) represents a leaf node in the recursive search tree, at which we either terminate the algorithm or traceback and try a different node. The target point of the base case is $\bold{p}^* = (\Bone^{*}, \Btwo^{*})$ only if its entropy $\Entropy^{*}$ is (i) minimal over all algorithmic history and (ii) is locally minimal compared with points $\bold{p}$ within the neighborhood
\begin{equation}
	 \mathcal{B}_h(\bold{p}) = \lbrace \bold{p}^* \in \mathbb{Z}_{+}^2 \mid \| \bold{p} - \bold{p}^* \|_{\infty} \leq h \rbrace \ ,
	 \label{eq:neighbors}
\end{equation}
where $h$ is a user-defined parameter that controls the size of the subproblem.

In Phase I, we perform MCMC at the trivial bipartite partition to create a reference description length. Then, starting from an initial state in which each node belongs to its own group, we apply an \texttt{Agglomerative-Merge} algorithm (also summarized in the main text) to reach a partition at $(\Bone^{\text{init}}, \Btwo^{\text{init}})$, where $\min\left(\Bone^{\text{init}}, \Btwo^{\text{init}} \right) = \lfloor \sqrt{2 E}/2 \rfloor$. The algorithm works as follows. In each sweep, we attempt $n_m$ block changes according to Eq.~\eqref{eq:proposal_moves} for each block. These proposal moves are not uniformly random, but are instead based on the current block structure, treating the edge count matrix $\bie$ as the adjacency matrix of a multigraph so that blocks can be thought of as nodes in this higher-level representation. Potential merges of blocks are then ranked according to increasing $\Delta \Entropy$, and exactly $\lceil B(1-\sigma^{-1}) \rceil$ block merges are performed in that order, in each sweep. To minimize the impact of bad merges done in the earlier steps, at the end of each sweep we apply MCMC algorithm described in the main text at zero temperature, allowing block changes that strictly decrease the entropy. This algorithm has an overall complexity of $\mathcal{O}\left( E \ln^2 N \right)$~\cite{peixotoEfficientMonteCarlo2014}, which is dwarfed when compared with the MCMC calculation. Note that in Sec.~\ref{sec:inference_algm}, we perform the same agglomerative merge algorithm right before the MCMC inference but merge blocks to a specific number of groups $\left( \Bone, \Btwo \right)$, rather than to a threshold given by $\min\left(\Bone^{\text{init}}, \Btwo^{\text{init}} \right) = \lfloor \sqrt{2 E}/2 \rfloor$. 

Phase II is the core recursive algorithm. Starting from the $\left(\Bone^{\text{init}}, \Btwo^{\text{init}} \right)$ partition, we check whether it is a local minimum in the description length landscape, where the radius of the local neighborhood is $h$, as defined in Eq.~\eqref{eq:neighbors}. If the current point is indeed a local minimum, the algorithm terminates. If it is not, the algorithm finds another candidate point in the $\left ( \bone, \btwo \right)$ grid by calling the \texttt{Rand-Merge} routine, which works by proposing many ways in which pairs of blocks could be merged, and then choosing the best merge. In particular, \texttt{Rand-Merge} proposes, for each block $r$, $n_m$ other blocks $s$ to which $r$ could be merged, selected uniformly at random.  From among those candidate merges, we choose the pair of blocks $r, s$ with the smallest relative entropy deviation $\delta_{r \sim s} = (\Entropy_{r \sim s} - \Entropy_{\text{ref}}) / \Entropy_{\text{ref}}$. Here, $\Entropy_{\text{ref}}$ is the minimum of all MCMC-calculated entropies explored globally and $\Entropy_{r \sim s}$ is the entropy that would result from a hypothetical merging of blocks $r$ and $s$.

\begin{figure}[b]
	\hrule height \phaserulewidth 
	\smallskip 
	\begin{algorithmic}[1]
		\Statex \textbf{Input}: Network $G=(V, E)$ with adjacency matrix $\biA$ and the partition $\bib^\text{0}$ which expresses each node belongs to which type.
		\Statex \textbf{Parameters} (we used the default values throughout this paper unless otherwise noted):
		\begin{itemize}
		\setlength\itemsep{-0.1em}
			\item $n_m=10$ number of merges attempted for each block
			\item $\sigma = 1.01$ greediness of agglomerative merges
			\item $\alpha=0.9$ adaptive parameter in case of overshooting
			\item $c=3$ trade-off parameter to determine $\Delta_0$ 
			\item $h=2$ neighborhood size

		\end{itemize}
		\Statex \textbf{Output}: Memoization table $\Xi$, which includes the MDL point.
		\phase{Initialization}
          \State $\Delta_0 \gets 1$
          \State Compute entropy $\Entropy^\text{0}$ for the trivial partition $\bib^\text{0}$
          \State Initiate memoization table $\Xi[1, 1] \gets (\bib^\text{0}, \Entropy^\text{0})$
          \State $(\Bone^{\text{init}}, \Btwo^{\text{init}}) \gets \textsc{Agglomerative-Merge}(\biA, n_m)$
		\phase{Dynamic Programming}
          \State Run $\textsc{Adaptive\_Search}(\Bone^{\text{init}}, \Btwo^{\text{init}})$
	\end{algorithmic}
	\hrule height \phaserulewidth
	\caption{Pseudocode for the search of the minimal entropy (or description length) point on the 2D landscape. The function \textsc{Adaptive\_Search} and its dependency \textsc{Local-Minimum\_Check} are described in Algm.~\ref{alg:helper}.}
	\label{alg:main}
\end{figure}

At this point, the algorithm gains its efficiency from avoiding calling the costly MCMC routine while still moving toward a local minimum in the $(\Bone, \Btwo)$ plane. To do so requires that we accept the merge and entropy change $\delta^* = \min\left( \lbrace \delta_{r \sim s}\rbrace \right)$ from \texttt{Rand-Merge} without pausing to re-fit the model using MCMC at the new $(\Bone, \Btwo)$. If this process is repeated, the entropy after accumulating merges will deviate more and more from the optimal entropy, were we to re-fit the model using MCMC at the current $(\Bone, \Btwo)$. We therefore balance speed and efficiency by introducing a parameter that forces a full MCMC fit only when the accumulated entropy from repeated merges becomes intolerable. Let $0 < \Delta_0 < 1$ such that when the a block merge does not deviate from the entropy too much ($\delta^* < \Delta_0$), we accept the merge and attempt the next successive merges. Otherwise, we seek to terminate the algorithm by calling \texttt{Local-Minimum\_Check} again at the current $(\Bone, \Btwo)$. 

The key to efficiency is that computing the approximated partition by block merges from an optimized partition is faster than finding it from scratch. Note that when the state of the algorithm is far from a local minimum, $\delta^* = \min\left( \lbrace \delta_{r \sim s}\rbrace \right)$ is typically small and negative, meaning that a large number of merges can often be performed before a full MCMC is required. Thus, choosing $\Delta_0$ is important. If we choose a large $\Delta_0$, the algorithm can overshoot the local minimum, requiring it to only gradually rediscover that minimum by inspecting many neighboring points. On the other hand, if we choose a small $\Delta_0$, there will be a larger number of MCMC calculations, which we also want to avoid. To this end, we determine $\Delta_0$ from the data on-the-fly during the \texttt{Adaptive\_Search} step. Namely, $\Delta_0$ is the first outlier $\delta^*$ based on the Interquartile Rule,
\begin{equation}\label{eq:iqr}
    \delta^* > c \text{IQR}(\lbrace \delta^* \rbrace) + Q_3(\lbrace \delta^* \rbrace) \ ,
\end{equation}
where $\lbrace \cdot \rbrace$ collects the $\delta^*$'s at earlier sweeps and the IQR is the interquartile range, being equal to the difference between $75^\text{th}$ and $25^\text{th}$ percentiles.  However, with this choice, we may still overshoot. In such cases, we reduce $\Delta_0$ by a factor $\alpha$ and relocate our attention to the $(\Bone^*, \Btwo^*)$ whose entropy $\Entropy^*$ is minimal so far, and then call \texttt{Local-Minimum\_Check}. During the neighborhood check, if we find an even better point nearby, we will relocate the tip to that point, and continue with the \texttt{Adaptive\_Search} step. The algorithm ends if a local minimum is found.

Because of our dynamic programming approach, the time complexity of the total algorithm cannot be computed directly from the recursion, nor do we know the exact number of subproblems (local searches using MCMC) that the algorithm will need to call. Indeed, as we found for synthetic networks near the detectability limit, and for networks near transitions in the resolution limit, the $(\Bone, \Btwo)$-optimization landscape becomes degenerate and multimodal, making a general algorithm complexity result hopeless. 

Nevertheless, the time complexity of the search algorithm scales with the number of MCMC calculations. Heuristic arguments suggest the number of MCMC calculations should be on the order of $\mathcal{O}(m h^2)$, where $m$ is the number of times that the most expensive for-loop [line 11 of \texttt{Local-Minimum\_Check}] is called. However, even this is an approximation, due to the fact that, at times, a local-minimum check reveals a point within the $h$-neighborhood that is better than the point currently being checked. In this way, subproblems may overlap, making the total cost somewhat cheaper. Empirically, for most networks in Table~\ref{table:benchmark}, $m < 3$.

\begin{figure}[b]
	\hrule height \phaserulewidth 
	\begin{algorithmic}[1]
		\Statex
        \Function{Adaptive\_Search}{$\Bone, \Btwo$} 
          \If{$\textsc{Local-Minimum\_Check}(\Bone, \Btwo)$}
            \State \Return{$\Xi$}  \Comment{Phase II terminates}
          \Else
            \State $\Delta \Entropy \gets 0$ 
            \State Update $\Entropy_{\text{ref}}$ to current MDL
            \While{$\Delta \Entropy < \Delta_0 \times \Entropy_{\text{ref}}$}
              \If{$\Bone \times \Btwo = 1$}
                \State \textbf{break}
              \Else
                \State $ r, s, \min({\lbrace \delta_{r \sim s} \rbrace}) \gets \textsc{Rand-Merge}(\bie, n_m)$
                \State $\Delta \Entropy \gets \min({\lbrace \delta_{r \sim s} \rbrace}) \times S_{\text{ref}}$
                \If{$\Delta_0 = 1$}
                  \State Update $\Delta_0$ if Eq.~\eqref{eq:iqr} is $\texttt{True}$
                  \State \textbf{break}
                \EndIf
                \State Update $\Bone, \Btwo$ from merged block pair $r, s$
                \State Update $\bie$ accordingly
              \EndIf
            \EndWhile
            \State \Return{$\textsc{Adaptive\_Search}(\Bone, \Btwo)$}
          \EndIf
        \EndFunction
		\Statex
	\end{algorithmic}
	\begin{algorithmic}[1]
		\Function{Local-Minimum\_Check}{$\Bone, \Btwo$}
          \State Compute entropy $\Entropy$ at $(\Bone, \Btwo)$ via Eq.~\eqref{eq:full_posterior}
          \State $\Xi[\Bone, \Btwo] \gets (\bib, \Entropy)$, where $\bib$ is the optimal partition
          \If{$\Entropy > \Entropy^{\text{0}}$}
            \State \Return $\texttt{False}$
          \EndIf
          \If{$\Entropy > \text{current MDL}$}  \Comment{Overshooting}
            \State $\Delta_0 \gets \alpha \Delta_0$
            \State Update $\Bone, \Btwo$ to which that give current MDL
          \EndIf
          \For{$\text{point } \bold{p} \in \mathcal{B}_h\left( \Bone, \Btwo \right)$}  \Comment{Refer Eq.~\eqref{eq:neighbors}}
            \State Compute entropy $\Entropy^{\bold{p}}$ via Eq.~\eqref{eq:full_posterior}
            \State $\Xi[\bold{p}_1, \bold{p}_2] \gets (\bib^\bold{p}, \Entropy^\bold{p})$
          \EndFor
          \If{$\Entropy > \text{current MDL}$}
            \State Update $\Bone, \Btwo$ to which that give current MDL
            \State \Return \texttt{False}
          \Else  
            \State $\Xi[\Bone, \Btwo] \gets (\bib, \Entropy)$
            \State \Return \texttt{True}
          \EndIf
		\EndFunction
	\end{algorithmic}
	\hrule height \phaserulewidth
	\caption{Pseudocode for the subroutines used in Algm.~\ref{alg:main}. Here, MDL is equivalent to the minimal MCMC-calculated entropy.}
	\label{alg:helper}
\end{figure}

     % Appendix A
     %
    %%%
   %%%%%
  %%%%%%%
    %%%
    %%%
\section{Prior for edge counts in the hierarchical biSBM}\label{appendix:hSBM}
    %%%    
    %%%   
  %%%%%%%
   %%%%%
    %%%
     %
     % Appendix B
In this appendix, we provide the prior for edge counts in the hierarchical bipartite SBM, corresponding to Eqs.~\eqref{eq:biclique_lower_multigraph} and Eq.~\eqref{eq:biclique_topmost_multigraph}. We begin with the flat SBM, whose prior for edge counts is,
\begin{equation} 
    P\left( \bie | \bib \right) = \multiset{\textmultiset{B}{2}}{E}^{-1} \ .	
    \label{eq:uniform_prior}
\end{equation}
In the hSBM, it might seem as if $P\left( \bie\mid \bib \right)$ should be written as a product of SBM likelihoods by repeatedly reusing Eq.~\eqref{eq:sbm_likelihood} at each additional level. However, at higher levels, the networks are multigraphs, and the SBM likelihood does not generate multigraphs uniformly because it is based on a uniform generation of configurations (i.e., $\Omega\left({\bik, \bie, \bib}\right)$) \cite{peixotoNonparametricBayesianInference2017}. Therefore, the correct way to build up the product is to directly count the number of multigraphs at each higher level using the dense ensemble~\cite{peixotoEntropyStochasticBlockmodel2012}, with each network instance occurring with the same probability.

\begin{table}[b]
\caption{List of model likelihood and prior functions, which contribute to the overall posterior probability function of different model variants used in this paper.}
\begin{tabular}{r | c | c | c | c | c}
	 \hline\hline
	{\bf Model variant} & SBM  & biSBM & \multicolumn{3}{c}{hSBM}  \\ \hline
	{\bf Hierarchy level} & \multicolumn{2}{c|}{$-$} & $0$ & $1, \dots, L-1$ & $L$ \\ \hline
	$P\left(\bie \mid \bib\right)$ & Eq.~\eqref{eq:uniform_prior} & Eq.~\eqref{eq:uniform_bipartite_prior} & $-$ & $-$ & Eq.~\eqref{eq:uniform_prior} \\ \hline
	$P\left(\biA \mid \bik, \bie, \bib \right)$ & \multicolumn{2}{c|}{Eq.~\eqref{eq:sbm_likelihood}} & Eq.~\eqref{eq:sbm_likelihood} & \multicolumn{2}{c}{Eq.~\eqref{eq:hi_edge_count_1}} \\ \hline
	$P\left(\bib\right)$ & \multicolumn{2}{c|}{Eq.~\eqref{eq:partitionprior}} & Eq.~\eqref{eq:partitionprior} & \multicolumn{2}{c}{Eq.~\eqref{eq:hi_edge_count_2}} \\ \hline
	$P\left(\bik \mid \bie, \bib\right)$ & \multicolumn{2}{c|}{Eq.~\eqref{eq:prior_deg}} & Eq.~\eqref{eq:prior_deg} & $-$ & $-$ \\ \hline
    Dense ensemble? & \multicolumn{2}{c|}{no} & no & \multicolumn{2}{c}{yes} \\
	 \hline\hline
\end{tabular}\label{table:parts_of_posterior}
\end{table}

Assuming that we have built $L$ higher-level models, the prior for edge counts between groups can be rewritten as
\begin{equation}
    P_{\text{hier}}\left( \bie \mid  \bib \right) = \prod_{l=1}^L P\left( \bie_l \mid  \bie_{l+1}, \bib_l \right) P\left(\bib_l \right) \ ,
    \label{eq:hi_edge_count}
\end{equation}
where
\begin{equation}\label{eq:hi_edge_count_1}
    P\left( \bie_l \mid  \bie_{l+1}, \bib_l \right) = \prod_{r<s} \multiset{n_r^l n_s^l}{e_{rs}^{l+1}}^{-1},
\end{equation}
and 
\begin{equation}\label{eq:hi_edge_count_2}
    P\left( \bib_l \right) = \frac{\prod_{r} n_r^{l+1}!}{B_l!} \binom{B_l - 1}{B_{l+1} - 1}^{-1} \frac{1}{B_l} \ .
\end{equation}
At the highest level ($l=L$), $\bie_L$ denotes a single-node multigraph with $E$ self-loops. Because there is no further block structure, we enforce $P\left( \bie_L \mid  \bie_{L+1}, \bib_l \right) = P\left( \bie_L \mid  \bib_l \right)$ and assume that the block is generated by a uniform prior, and reuse Eq.~\eqref{eq:uniform_prior}.

One peculiar consequence of forcing the hSBM, implemented in \texttt{graph-tool}, to consider only type-specific blocks is that even when a network has no statistically justifiable structure, the hSBM finds the trivial bipartite partition {\it and then} builds a final hierarchical level on that trivial bipartite partition. In other words, it cannot help  but find a single group at the topmost level. This explains the otherwise perplexing distribution of model description lengths shown in Fig.~\ref{fig:empirical}a: both the SBM and hSBM find the trivial partition, but this partition is more costly to express via the hSBM due to our having forced it to respect the network's bipartite structure.  Table~\ref{table:parts_of_posterior} summarizes all model likelihood and prior functions pertinent to this paper, for reference.
     % Appendix B
     %
    %%%
   %%%%%
  %%%%%%%
    %%%
    %%% 
%\clearpage
\bibliography{biSBM}
\appendix 

\end{document}